\documentclass[useAMS,usenatbib]{mn2e}
\input psfig.sty

\newcommand {\apj} {ApJ}
\newcommand {\apjl} {ApJL}
\newcommand {\apjs} {ApJS}
\newcommand {\mnras} {MNRAS}

\newcommand {\aap} {A\&A}
\newcommand {\aj} {AJ}

\newcommand {\etal} {et~al.~}
\def \spose#1{\hbox  to 0pt{#1\hss}}  
\newcommand {\lta} {\mathrel{\spose{\lower 3pt\hbox{$\sim$}}\raise  2.0pt\hbox{$<$}}}
\newcommand {\gta} {\mathrel{\spose{\lower  3pt\hbox{$\sim$}}\raise 2.0pt\hbox{$>$}}}

\newcommand {\kms} {\ifmmode  \,\rm km\,s^{-1} \else $\,\rm km\,s^{-1}  $ \fi }
\newcommand {\kpc} {\ifmmode  {\rm kpc}  \else ${\rm  kpc}$ \fi  }  
\newcommand {\Msun} {\ifmmode M_{\odot} \else $M_{\odot}$ \fi} 
\newcommand {\hMsun} {\ifmmode h^{-1}\,\rm M_{\odot} \else $h^{-1}\,\rm M_{\odot}$ \fi}
\newcommand {\Gyr} {\ifmmode  {\rm Gyr}  \else ${\rm  Gyr}$ \fi  }  

\newcommand {\LCDM} {\ifmmode \Lambda{\rm CDM} \else $\Lambda{\rm CDM}$ \fi}
\newcommand {\fbar} {\ifmmode f_{\rm bar} \else $f_{\rm bar}$ \fi} 
\def \Deltavir {\ifmmode \Delta_{\rm vir} \else $\Delta_{\rm vir}$ \fi}
\def \OmegaL {\ifmmode \Omega_{\rm \Lambda} \else $\Omega_{\rm \Lambda}$\fi} 

\newcommand {\Rvir} {\ifmmode R_{\rm vir} \else $R_{\rm vir}$ \fi}
\newcommand {\Vvir} {\ifmmode V_{\rm  vir} \else  $V_{\rm vir}$  \fi} 
\newcommand {\Mvir} {\ifmmode M_{\rm  vir} \else $M_{\rm  vir}$ \fi}  
\newcommand {\Jvir} {\ifmmode J_{\rm vir} \else $J_{\rm vir}$ \fi} 
\newcommand {\Evir} {\ifmmode E_{\rm vir} \else $E_{\rm vir}$ \fi} 
\newcommand {\lam} {\ifmmode \lambda  \else $\lambda$ \fi} 
\def \cvir {\ifmmode c_{\rm vir} \else $c_{\rm vir}$ \fi} 

\newcommand {\lamgal} {\ifmmode \lambda_{\rm gal}  \else $\lambda_{\rm gal}$ \fi} 
\newcommand {\lamstar} {\ifmmode \lambda_{\rm star}  \else $\lambda_{\rm star}$ \fi} 
\newcommand {\lamcold} {\ifmmode \lambda_{\rm cold}  \else $\lambda_{\rm cold}$ \fi}

\newcommand{\tcmin}{\ifmmode t_{\rm c,min} \else $t_{\rm c,min}$ \fi}
\newcommand{\tcmax}{\ifmmode t_{\rm c,max} \else $t_{\rm c,max}$ \fi}
\newcommand{\Zhot}{\ifmmode Z_{\rm hot} \else $Z_{\rm hot}$ \fi}

\newcommand {\mgal}    {\ifmmode m_{\rm gal}    \else $m_{\rm gal}$ \fi} 
\newcommand {\mstar}    {\ifmmode m_{\rm star}    \else $m_{\rm star}$ \fi} 
\newcommand {\mcold}    {\ifmmode m_{\rm cold}    \else $m_{\rm cold}$ \fi} 
\newcommand {\Mgal}  {\ifmmode M_{\rm gal}  \else $M_{\rm gal}$ \fi}
\newcommand {\Mstar}    {\ifmmode M_{\rm star}    \else $M_{\rm star}$ \fi} 
\newcommand {\Mcold}    {\ifmmode M_{\rm cold}    \else $M_{\rm cold}$ \fi}

\newcommand {\jgal} {\ifmmode j_{\rm gal} \else $j_{\rm gal}$ \fi}  
\newcommand {\Jgal} {\ifmmode J_{\rm gal} \else $J_{\rm gal}$ \fi}  
\newcommand {\jstar} {\ifmmode j_{\rm star} \else $j_{\rm star}$ \fi}  
\newcommand {\Jstar} {\ifmmode J_{\rm star} \else $J_{\rm star}$ \fi}  
\newcommand {\jcold} {\ifmmode j_{\rm cold} \else $j_{\rm cold}$ \fi}  
\newcommand {\Jcold} {\ifmmode J_{\rm cold} \else $J_{\rm cold}$ \fi}

\newcommand {\Vrot} {\ifmmode  V_{\rm rot} \else $V_{\rm rot}$ \fi} 
\def \Vmax {\ifmmode V_{\rm  max} \else  $V_{\rm max}$  \fi} 
\def \Vmaxh {\ifmmode V_{\rm  max,h} \else  $V_{\rm max,h}$  \fi} 

\title[Feedback and Disk Galaxy Scaling Relations]
      {The Impact of Feedback on Disk Galaxy Scaling Relations}
    
\author[Dutton \& van den Bosch]
       {Aaron A. Dutton$^{1,2}$\thanks{dutton@ucolick.org} \& 
         Frank C. van den Bosch$^3$\\
      $^1$UCO/Lick Observatory and Department of Astronomy and Astrophysics, University of California, Santa Cruz, CA 95064, USA\\
      $^2$Department of Physics, Swiss Federal Institute of Technology
          (ETH Z\"urich), CH-8093 Z\"urich, Switzerland \\
      $^3$Max-Planck-Institut f\"ur Astronomie, K\"onigstuhl 17, 69117
          Heidelberg, Germany}

\begin{document}
             
\date{submitted to MNRAS}
             
\pagerange{\pageref{firstpage}--\pageref{lastpage}}\pubyear{2009}

\maketitle           

\label{firstpage}


\begin{abstract}
  We use  a disk galaxy evolution  model to investigate  the impact of
  mass outflows  (a.k.a.  feedback) on disk  galaxy scaling relations,
  mass fractions and spin parameters. Our model follows the accretion,
  cooling, star formation and ejection of baryonic mass inside growing
  dark matter  haloes, with cosmologically  motivated angular momentum
  distributions, and dark matter halo structure.
%
  Models without feedback produce disks that are too small, too gas
  poor, and which rotate too fast.  Feedback reduces the galaxy formation
  efficiency, $\epsilon_{\rm GF}$, (defined as the fraction of the
  universally available baryons that end up as stars and cold gas in a
  given galaxy), resulting in larger disks with higher gas fractions
  and lower rotation velocities.  Models with feedback can reproduce
  the zero points of the scaling relations between rotation velocity,
  stellar mass and disk size, but only in the absence of adiabatic
  contraction.  Our feedback mechanism is maximally efficient in
  expelling mass, but our successful models require 25\% of the SN
  energy, or 100\% of the SN momentum, to drive an outflow.  It
  remains to be seen whether such high efficiencies are realistic or
  not.
%
  Our  energy and  momentum  driven wind  models  result in  different
  slopes of various scaling  relations.  Energy driven winds result in
  steeper slopes  to the galaxy mass  - halo mass, and  stellar mass -
  halo mass relations, a shallower  slope to the galaxy size - stellar
  mass  relation  at  $z=0$, and  a  steeper  slope  to the  cold  gas
  metallicity -  stellar mass  relation at $z\simeq  2$.  Observations
  favor the  energy driven wind  at stellar masses below  $\Mstar \lta
  10^{10.5}\Msun$, but the momentum driven wind model at high masses.
%
  The ratio  between the specific  angular momentum of the  baryons to
  that of the  halo, $(\jgal/\mgal)$, is not unity  in our models with
  inflow and outflow.   Yet this is the standard  assumption in models
  of disk formation.  Above a halo mass of $\Mvir\simeq 10^{12}\Msun$,
  cooling   becomes  increasingly   inefficient,   which  results   in
  $(\jgal/\mgal)$ decreasing with increasing  halo mass.  Below a halo
  mass of  $\Mvir \simeq 10^{12}\Msun$,  feedback becomes increasingly
  efficient.   Feedback  preferentially  ejects low  angular  momentum
  material  because  star  formation  is  more  efficient  at  smaller
  galactic  radii,   and  at   higher  redshifts.   This   results  in
  $(\jgal/\mgal)$ increasing  with decreasing halo  mass.  This effect
  helps to  resolve the discrepancy  between the high  spin parameters
  observed for  dwarf galaxies with the low  spin parameters predicted
  from \LCDM.

\end{abstract}

\begin{keywords}
  galaxies: formation --  galaxies fundamental parameters -- galaxies:
  haloes --  galaxies: kinematics and dynamics --  galaxies: spiral --
  galaxies: structure 
\end{keywords}

\setcounter{footnote}{1}


\section{Introduction}
\label{sec:intro}
Understanding the origin  and nature of galaxy scaling  relations is a
fundamental quest  of any successful theory of  galaxy formation.  The
success  of a  particular  theory will  be  judged by  its ability  to
reproduce  the slope,  scatter, and  zero-point of  any  robust galaxy
scaling relation.   Of particular  interest are the  scaling relations
between  size ($R$),  luminosity  (or stellar  mass,  $M$) ($L$),  and
velocity ($V$), as these parameters  are related to each other via the
virial theorem.   For early type galaxies, these  three parameters are
indeed  coupled, resulting  in a  two-dimensional plane  known  as the
Fundamental Plane (FP: Dressler \etal 1987; Djorgovski \& Davis 1987).
However,  for  late-type   galaxies,  the  relation  between  rotation
velocity and luminosity, known as the Tully-Fisher relation (TF: Tully
\& Fisher  1977) is independent  of galaxy size or  surface brightness
(e.g. Courteau \& Rix 1999;  Courteau \etal 2007; Pizagno \etal 2007).
Understanding the  origin of  this surface brightness  independence is
likely the key  to understanding the small scatter of  the TF, and may
even explain the origin of the zero point of the TF relation.

The slopes  of the  $VLR$ relations for  disk galaxies can  be broadly
understood   with  galaxy   formation  models   that   include  virial
equilibrium  after dissipation-less  collapse  of quasi-spherical  cold
dark matter  (CDM) halos and angular momentum  conservation (e.g.  Mo,
Mao, \&  White 1998;  van den Bosch  1998, 2000; Navarro  \& Steinmetz
2000; Firmani \& Avila-Reese 2000; Dutton \etal 2007).

However,  reproducing these  scaling relations  in detail  has  been a
problem for  galaxy formation  models.  To date,  no (self-consistent)
CDM-based model of galaxy formation can {\it simultaneously} match the
the zero points  of the TF relation, galaxy  sizes, and the luminosity
(or  stellar mass)  function (e.g.   Benson \etal  2003;  Dutton \etal
2007).  Models which match the zero point of the TF relation, do so by
making the assumption that, $\Vrot$, the observed rotation velocity is
equal to,  $\Vvir$, the circular  velocity at the virial  radius (e.g.
Somerville  \&  Primack  1999),  or,  $\Vmaxh$,  the  maximum  circular
velocity of  the halo prior  to galaxy formation, (e.g.   Croton \etal
2006).  For typical galaxy  mass dark matter haloes $\Vmaxh/\Vvir\simeq
1.1$ (e.g.  Bullock \etal 2001a),  so these two assumptions are almost
equivalent.   Other  more observational  approaches  also support  the
conclusion  that $\Vrot\simeq  \Vmaxh$ (Eke  \etal 2006;  Blanton \etal
2008).   The  problem  for   galaxy  formation  theory  is  that  both
cosmological simulations and analytic  models of disk galaxy formation
that take into account the  self-gravity of the baryons and the effect
of halo  contraction (Blumenthal \etal  1986) find that  $\Vrot \simeq
1.8 \Vvir$ (Navarro  \& Steinmetz 2000; Abadi \etal  2003; Dutton \etal
2007; Governato \etal 2007).

Due to the almost  one-to-one correlation between $\Vmaxh$ and $\Mvir$,
and  the correlation between  $\Vrot$ and  baryonic $\Mgal$  mass (the
baryonic  TF   relation  e.g.   McGaugh  \etal   2000)  an  equivalent
constraint to the ratio between $\Vrot$ and $\Vmaxh$ is the galaxy mass
fraction:  $\mgal =  \Mgal/\Mvir$. Observations  of halo  masses using
weak lensing  (Hoekstra \etal 2005;  Mandelbaum \etal 2006)  find that
for late  type galaxies the  maximum $\mgal \simeq 0.33  f_{\rm bar}$,
where  $\fbar=\Omega_{\rm  b}/\Omega_{\rm  m,0}  \simeq 0.17$  is  the
cosmic baryon mass fraction  at redshift zero.  Similarly low galaxy
formation efficiencies are obtained  by methods that match the stellar
mass function with the dark halo mass function (e.g.  Yang \etal 2007;
Conroy \& Wechsler 2008).

A partial  explanation for the surface brightness  independence of the
TF relation, or equivalently the weak correlation between residuals of
the $VL$ and $RL$ relations,  is that observationally (e.g. McGaugh \&
de Blok  1997) gas fractions  correlate with surface  brightness, with
higher gas fractions in lower surface brightness galaxies. Since lower
surface density  disks are expected to  rotate more slowly  at a given
baryonic mass, the larger gas fractions shift these galaxies to lower
stellar masses  and hence  lower luminosities (Firmani  \& Avila-Reese
2000;  van  den  Bosch  2000).  However, this  solution  is  not  very
effective for the high  surface brightness, low gas fraction galaxies.
Dutton \etal (2007) showed that  a reasonable amount of scatter in the
stellar mass-to-light  ratios helps to reduce  the correlation between
the residuals  of the $VL$  and $RL$ relations further.   Gnedin \etal
(2007)  proposed a  correlation between  disk mass  fraction  and disk
surface density to explain the  lack of a correlation between the $VM$
and $RM$ relations.

Thus understanding the physical  mechanisms that determine galaxy mass
fractions is  fundamental to  our understanding of  the origin  of the
$VLR$ relations  of disk  galaxies.  In the  standard picture  of disk
galaxy formation,  gas that enters the  halo gets shock  heated to the
virial  temperature, and  then cools  radiatively (Fall  \& Efstathiou
1980).  Thus in  order to produce low galaxy  mass fractions, either a
significant fraction of the gas has to be prevented from cooling, or a
significant fraction  must subsequently be  ejected from the  disk and
halo.  The latest hydrodynamical simulations indicate that rather than
accreting gas via a cooling flow, below a critical halo mass of $\Mvir
\simeq 10^{12}\Msun$  gas does  not shock as  it enters the  halo, and
instead accretes straight onto the  galaxy in a cold flow (Birnboim \&
Dekel 2003; Keres  \etal 2005).  In this scenario,  essentially all of
the baryons accrete  onto the galaxy, and the  problem of stopping the
baryons  from cooling  becomes one  of  stopping the  cold flows  from
forming.  Thus  in the absence of  some kind of  pre-heating (e.g.  Mo
\etal 2005)  which would shut down  the cold flows,  mass outflows are
required in order to produce the low galaxy mass fractions observed in
the Universe today.

In order to investigate how mass outflows (a.k.a. feedback) determines
galaxy mass fractions, and the  impact this has on disk galaxy scaling
relations,  we use  an updated  version of  the disk  galaxy evolution
models presented in van den  Bosch (2001; 2002a).  In these models the
input parameters  are the  concentration and spin  of the  dark matter
halo.   The galaxy  spin parameter,  galaxy mass  fraction and  gas to
stellar mass ratio are  collectively determined by the efficiencies of
cooling, star formation, and feedback.

The main differences  with respect to the van  den Bosch (2002a) models
are:
\begin{enumerate}
\item  We  use  cosmologically  motivated  specific  angular  momentum
  distributions (AMDs)  of the  halo gas rather  than shells  in solid
  body rotation;
\item We  consider a  star formation recipe  based on  dense molecular
  gas,  rather  than  on  total  gas  with  a  Toomre  star  formation
  threshold;
\item We include scatter in  halo concentration which we relate to the
  mass accretion history (MAH);
\item We explore  two different feedback models: one  based on kinetic
  energy conservation, the other based on momentum conservation.
\end{enumerate}

An important  aspect of this model is  that we do not  assume that the
baryonic disk has  an exponential density profile.  In  this model the
surface density profile of the baryonic disk is determined by detailed
conservation of angular momentum, starting from the AMDs of gas haloes
as found in cosmological  simulations.  The surface density profile of
the  stars  is  then   determined  by  the  relative  efficiencies  of
star-formation, outflows,  and inflows as a function  of radius.  This
allows  us to  self-consistently follow  the evolution  of  the radial
distributions of  gas and stars. In  future papers we  use this galaxy
formation  model to  investigate  the origin  of  disk galaxy  surface
density  profiles  (Dutton 2008)  and  the  evolution  of disk  galaxy
scaling relations.

This paper  is organized as  follows: In \S \ref{sec:dfm}  we describe
the disk galaxy  formation models; in \S \ref{sec:vlr}  we discuss the
effect of  feedback on  the $VMR$ relations;  in \S  \ref{sec:mgal} we
discuss the  impact of  feedback on galaxy  mass fractions  and galaxy
spin parameters;  in \S\ref{sec:mz} we discuss the  effect of feedback
on the mass-metallicity relation  at $z\simeq 2$; in \S \ref{sec:mdot}
we discuss  how galaxies  lose their gas;  and in \S  \ref{sec:sum} we
give a summary.


\section{Disk Galaxy Formation Models}
\label{sec:dfm}
The main assumptions that characterize the framework of our models are
the following: 
\begin{enumerate}
\item  Dark matter  haloes around  disk  galaxies grow  by the  smooth
  accretion of mass; 
\item The  baryons acquire the  same distribution of  specific angular
  momentum as the dark matter;
\item Gas that enters the halo is shock heated  to the virial temperature;
\item Gas  cools conserving its  specific angular momentum;
\item Star  formation occurs  according to a  Schmidt type law  on the
  dense molecular gas;
\item Supernova feedback re-heats some  of the cooled gas, ejecting it
  from the halo;
\item  The halo  contracts and  expands adiabatically  to  inflows and
  outflows, respectively;
\end{enumerate}

Each model galaxy  consists of five mass components:  dark matter, hot
halo gas, disk mass in stars,  disk mass in cold gas, and ejected gas.
The  dark matter  and the  hot gas  are assumed  to be  distributed in
spherical shells,  the disk mass  is assumed to be  in infinitesimally
thin annuli.  Throughout this paper we  refer to $R$ as radius, $t$ as
time (where $t=0$ is defined as the Big Bang) and $z$ as redshift.

For each  galaxy we set up  a radial grid with  200 bins quadratically
sampled from between 0.001 to 1 times the redshift zero virial radius,
and  we follow the  formation and  evolution of  the galaxy  using 400
redshift steps  quadratically sampled from $z=10$ to  $z=0$.  For each
time step  we compute  the changes in  the various mass  components in
each radial  bin.  The  prescriptions we use  are described  in detail
below.

\subsection{Limitations of the Model}
Before  we describe the  details of  the model,  we first  discuss the
relevance  of  our  model  to  understanding  galaxy  formation  in  a
hierarchical universe.

The assumption  of smooth mass accretion might  seem inconsistent with
the  hierarchical  merger picture  of  structure  formation  in a  \LCDM
universe.  However,  major mergers of stellar rich  galaxies are known
to destroy disks (e.g. Barnes  1992; Cox \etal 2006), so disk galaxies
are  unlikely to  form in  haloes  with recent  major mergers.   Minor
mergers are more common than major mergers, and are likely to play and
important role in the formation  of galaxy bulges, either directly, or
by triggering secular  processes.  Thus by not  including mergers our
model underestimates  bulge fractions.  However,  one of the  goals of
these models is to determine  how much of the structural properties of
disk galaxies can be accounted  for with the ``zeroth order'' scenario
of disk formation (smooth accretion and quiescent star formation).

Our  assumption about  the way  gas is  accreted into  galaxies  (by a
cooling  flow of shock  heated gas)  is likely  incorrect. Simulations
suggest  that disk  galaxies accrete  most of  their mass  though cold
flows, and that in the absence of extra heating or outflows the baryon
fraction  of galaxies  is close  to the  universal value  (Keres \etal
2005).  In  our model the gas  shock heats, but since  for haloes with
masses below $10^{12}\Msun$ the cooling  time is short compared to the
Hubble time,  essentially all  the gas that  enters the  halo accretes
onto  the  disk in  a  free fall  time.   Thus  although the  physical
mechanism  by  which galaxies  accrete  their  gas  in our  model  and
cosmological simulations  are different,  we expect that  the specific
angular momentum distribution  of cold flow baryons to  be the same as
that  of the  dark  matter, though  this  needs to  be verified  using
cosmological  simulations.

Our assumption about the efficiency at which gas is converted into
stars is necessarily over-simplified, but it is an improvement over
the majority of star formation recipes (which may be physically or
empirically motivated) used in semi-analytic models and
hydro-dynamical simulations, which are based on the density (or mass)
of the total cold gas (atomic plus molecular).  We assume an empirical
relation between the local star formation rate and the local density
of molecular gas.  We calculate the molecular gas fraction (as a
function of radius in the disk) using the empirical relation between
mid plane pressure and molecular gas fraction from Blitz \& Rosolowski
(2006).  A more realistic treatment of star formation would model the
formation and destruction of molecular gas in a physically motivated
way (e.g. Pelupessy, Papadopoulos, \& van der Werf 2006; Robertson \&
Kravtsov 2008)

In our feedback  model we only consider winds that  are able to escape
the halo, and  we assume that mass in such winds  is lost forever.  In
reality  mass that  escapes the  halo may  fall back  at  later times.
Furthermore, there  may be winds  that have enough  energy/momentum to
escape the disk  but not the halo.  The gas in  these winds could then
re-cool back onto  the disk producing a galactic  fountain. Since very
little is known about how feedback  works, and our main interest is to
determine how much mass can be  ejected from the disk and halo, rather
than  introducing additional  free parameters  to our  wind  model, we
assume the maximal mass loss model.

The assumption that the halo responds adiabatically to inflows and
outflows may not be correct.  When galaxies accrete their gas via a
smooth cooling flow the gas radiates away its energy.  However, when a
galaxy acquires its gas via cold flows, clumps of cold gas can
exchange energy with the halo via dynamical friction (e.g.  El Zant,
Shlosman, \& Hoffman 2001; El-Zant \etal 2004; Jardel \& Sellwood
2009), causing the halo to expand.  In both scenarios the natural
response of the halo to the deepening of the potential well due to the
condensation of baryons is to contract, but in the latter scenario,
the transfer of energy between baryons and dark matter will counter
this effect.  Processes internal to disks, such as bars, can also
transfer energy to the halo via dynamical friction, causing it to
expand (Weinberg \& Katz 2002; Holley-Bockelmann, Weinberg \& Katz
2005; Sellwood 2008).  Mass outflows can also result in halo
expansion, for example an adiabatic inflow followed by an
instantaneous outflow can result in net halo expansion (Navarro, Eke,
\& Frenk 1996; Gnedin \& Zhao 2002; Read \& Gilmore 2005). Thus in
order to asses how much impact the assumption of adiabatic inflow and
outflow has on the structural properties of the resulting galaxies, we
also consider models in which the halo does not respond to galaxy
formation.

\subsection{Dark Matter Haloes}
The backbone of  our galaxy formation model is the  growth of the dark
matter halo, which we model  by a smooth mass accretion history (MAH).
Van den  Bosch (2002b) and Wechsler  \etal (2002) have  shown that the
MAH  is essentially  a one  parameter family.   The MAH  from Wechsler
\etal (2002) is given by
\begin{equation}
\label{eq:MAH}
\Mvir(z) = \Mvir\, e^{-\alpha z},
\end{equation}
where $\Mvir$ is the redshift zero mass and $\alpha$ is related to
the epoch of formation via
\begin{equation}
\label{eq:acS}
\alpha=a_{\rm c} \,S.
\end{equation}
Here $a_{\rm  c} =(1+z_{\rm c})^{-1}$  is defined as the  scale factor
$a$ when the logarithmic slope  of the accretion rate falls below some
specified value, $S$. Following  Wechsler \etal (2002) we adopt $S=2$.
Before  we discuss  how  to compute  $a_{\rm  c}$ we  discuss how  the
structural properties of the halo depend on its mass.

In the standard spherical top-hat collapse model the 
virial radius, $\Rvir(z)$, of a halo of mass $\Mvir(z)$ at a redshift,
$z$, is given by
\begin{eqnarray}
\label{eq:RMz}
\left[\frac{\Rvir(z)}{h^{-1}\rm kpc}\right]\simeq162.6
\left[\frac{\Mvir(z)}{10^{12}h^{-1}\Msun}\right]^{\frac{1}{3}} 
\left[\frac{200}{\Delta_{\rm vir}(z)}\right]^{\frac{1}{3}}
\left[\frac{H_0}{H(z)}\right]^{\frac{2}{3}},
\end{eqnarray}
where  $h=H_0/100$, 
and $\Delta_{\rm vir}$ is the virial density, relative to the critical
density for  closure. We  use the fitting  formula of Bryan  \& Norman
(1998)
\begin{equation}
  \Delta_{\rm vir} = 18\pi^2 + 82x-39x^2, 
\end{equation}
with $x=\Omega_{\rm m}(z) -1$.
The evolution of the matter density is given by
\begin{equation}
\Omega_{\rm m}(z) = \Omega_{\rm m,0} (1+z)^3 \left[\frac{H(z)}{H_0}\right]^{-2}.
\end{equation}
and the evolution of the Hubble parameter is given by
\begin{equation}
H^2(z)=H_0^2\,[\Omega_{\Lambda} + (1-\Omega_{\Lambda}
- \Omega_{\rm m,0})(1+z)^2 + \Omega_{\rm m,0}(1+z)^3].
\end{equation}
The relation between the  virial velocity, $\Vvir$, and virial radius,
$\Rvir$ is given by
\begin{equation}
\Vvir = \sqrt{\frac{G \Mvir}{\Rvir}}
\end{equation} 

We assume that the density profile of the halo at each redshift is
given by an NFW (Navarro, Frenk, \& White 1997) profile
\begin{equation}
\frac{\rho(R)}{\rho_{\rm crit}} =  \frac{\delta_{\rm c}}{ (R/R_{\rm
    s}) (1+R/R_{\rm s})^{2}},
\end{equation}
where $R_{\rm s}$ is the radius where the slope of the density profile
is -2, the  so called scale radius, $\rho_{\rm  crit}$ is the critical
density of  the universe, and  $\delta_{\rm c}$ is  the characteristic
overdensity of  the halo. The  concentration parameter of the  halo is
defined as $c=\Rvir/R_{\rm s}$ and is related to $\delta_{\rm c}$ via
\begin{equation}
\delta_{\rm c} = \frac{\Deltavir}{3}\frac{c^3}{[ \ln(1+c) - c/(1+c) ]}.
\end{equation}

Following  Bullock  \etal  (2001a)  we assume  that  the  concentration
parameter evolves as
\begin{equation}
\label{eq:cz}
c(z) = K a(z) / a_{\rm c} = K \frac{1+z_{\rm c}}{1+z}
\end{equation}
with  a  minimum  value  of  $c(z)=K$,  corresponding  to  a  constant
concentration  during the rapid  accretion phase  of dark  halo growth
(Zhao \etal 2003).  Thus if we specify the redshift zero concentration
we can compute the collapse epoch via $a_{\rm c} = K/c$, and hence the
MAH through Eqs.~(\ref{eq:MAH}) \& (\ref{eq:acS}).

To compute the mean concentration for  a halo of a given mass at $z=0$
we use  the Bullock  \etal (2001a) model.   This model requires  us to
specify  the  cosmology $\OmegaL,  \Omega_{\rm  m,0}, \Omega_{\rm  b},
\sigma_8, n, h$, as well as 2 free parameters, $F$ and $K$.  We assume
the concentration is log-normally  distributed with median $c(M)$ from
the  Bullock \etal (2001a)  model and  scatter $\sigma_{\ln  c}$.  Our
adopted values for these parameters are given in \S \ref {sec:param}.

\subsection{Gas Cooling}
\label{sec:cooling}
At   each    time   step   a    shell   with   mass   $\Delta    M   =
\Mvir(t)-\Mvir(t-\Delta  t)$  virializes. A  fraction  $f_{\rm bar}  =
\Omega_{\rm b}/\Omega_{\rm m,0}$  of this mass  is in  baryons, and  is heated  to the
halo's virial temperature
\begin{equation}
T_{\rm vir} = \frac{1}{2} \frac{\mu\, m_{\rm p}}{k} \Vvir^2,
\end{equation}
where  $\mu\,m_{\rm  p}$  is  the   mass  per  particle,  and  $k$  is
Boltzmann's constant.  The  baryons dissipate energy radiatively, lose
pressure   support,  and   collapse  until   they   reach  centrifugal
equilibrium.   The time  scale  over  which this  occurs  is given  by
$t_c\equiv {\rm  max}[t_{\rm ff},t_{\rm cool}]$. Here  $t_{\rm ff}$ is
the free fall time defined as
\begin{equation}
t_{\rm ff} = \sqrt{ \frac{3\pi}{32 G \bar{\rho}} },
\end{equation}
with $\bar{\rho}$ the average halo density, and 
\begin{equation}
t_{\rm cool} = \frac{3}{2} \mu m_{\rm p} \frac{k T_{\rm
    vir}}{\rho_{\rm hot} \Lambda_{\rm N}(Z_{\rm hot})}
    \frac{\mu_{\rm e}^2}{\mu_{\rm e} -1}
\end{equation}
is the cooling  time. Here $\rho_{\rm hot}$ is the  density of the hot
gas,  $\mu_{\rm e}$  is  the  number of  particles  per electron,  and
$\Lambda_{\rm N}$  is the normalized  cooling function for a  gas with
metallicity $Z_{\rm  hot}$. For $\rho_{\rm hot}$ we  use $f_{\rm bar}$
times the density at the virial  radius at the time the gas enters the
halo.   For  $\Lambda_{\rm  N}$  we  use  the  collisional  ionization
equilibrium cooling functions of Sutherland \& Dopita (1993), assuming
a helium mass abundance of 0.25.

For each time step, we compute the range of times between which gas that
collapses  onto the  disk in  the  current time  interval, entered  the
halo.  We  label  these   times  as  $t_{\rm  c,min}(t)$  and  $t_{\rm
  c,max}(t)$.

\subsection{Angular Momentum Distribution}
The radius  at which the  cooled gas ends  up depends on  its specific
angular   momentum,  $j$.    Van  den   Bosch  (2001; 2002a)  assumed the
$j$-distribution to  be that  of a shell  in solid body  rotation. 
The angular momentum  of this shell can be  computed assuming that the spin
parameter, $\lambda$, is  constant between time  steps.  
The spin parameter is defined by
\begin{equation}
\lambda = \frac{\Jvir | \Evir |^{1/2}}{G\Mvir^{5/2}},
\end{equation}
where $\Mvir, \Jvir$, and $\Evir$ are the mass, total angular momentum
and  energy of  the halo,  respectively.  As  shown by  van  den Bosch
(2001)  this results in  density profiles  that are  more concentrated
than  exponential in  the center,  and also  that truncate  at shorter
radii than is observed in some disk galaxies.

Building on  the work  of Bullock \etal  (2001b), Sharma  \& Steinmetz
(2005,  hereafter SS05)  used  a series  of  non radiative  N-body/SPH
simulations in a $\Lambda$CDM cosmology to study the growth of angular
momentum in galaxy  systems.  They introduced a function  that is able
to describe,  with a single parameter, $\alpha$,  the specific angular
momentum distribution of the gas and dark matter in their simulations,
as well as that of exponential disks in NFW haloes:
\begin{equation}
\frac{M(<j)}{\Mvir}=\gamma(\alpha,\frac{j}{j_{\rm d}}), \;\; j_{\rm d} = \frac{\Jvir}{\Mvir} \frac{1}{\alpha}
\end{equation}
where $\gamma$ is the incomplete  gamma function.  SS05 found that the
distribution of $\alpha$ is 
log-normally   distributed  with   mean   $\log\alpha\simeq-0.05$  and
standard deviation in $\log\alpha\simeq0.11$.

In summary the  distribution of specific angular momentum  of the dark
matter  halo  and  hot gas  can  be  described  by two  parameters:  a
normalization   ($\lambda$)  and   a  shape   ($\alpha$).    Both  the
normalization and shape  parameters are log-normally distributed, with
significant scatter.  We assume that the spin and shape parameters are
uncorrelated, although Bullock \etal (2001b)  show that there may be a
weak correlation between $\lambda$ and $\alpha$. Furthermore we assume
that,  for a  given halo,  $\lambda$  and $\alpha$  are constant  with
redshift.  These assumptions  are made for simplicity, and  need to be
tested with cosmological simulations.

In order to compute the amount of mass, with a given specific angular
momentum, $j$, that has collapsed onto the disk in each time step, $M_{\rm
  c}$, we take the difference between the distributions of specific
angular momentum of the halos at times \tcmin and \tcmax ($\tcmin$ and
$\tcmax$ are defined in \S \ref{sec:cooling} above);
\begin{equation}
\label{eq:mcoll}
M_{\rm c}(<j)(t) = f_{\rm bar}[ \Mvir(<j)(\tcmax) - \Mvir(<j)(\tcmin)].
\end{equation}

\subsection{Conservation of Angular Momentum and Halo Contraction}
\label{sec:am}
In order to  compute the radius, $R$, at  which material with specific
angular momentum, $j$, ends up  we assume specific angular momentum is
conserved, i.e. one should solve
\begin{equation}
j = R\, V_{\rm circ}(R) 
\end{equation}
for $R$.   Here $V_{\rm  circ}$ is the  total circular  velocity (from
stars, cold gas, hot gas, and dark matter).

As the galaxy grows over time  the circular velocity at a given radius
increases. Thus  to conserve specific angular  momentum, material that
settled at radius $R$ would need  to drift to smaller radii over time.
Given that  the gas and  stars effect the circular  velocity, strictly
conserving the specific  angular momentum of the baryons  over time is
difficult to implement numerically.  To get around this problem we use
specific  angular momentum,  $j$,  rather than  radius  as our  radial
coordinate.

Under   the   simplifying    assumption   that   $V_{\rm   circ}=   [G
M(<R)/R]^{1/2}$,  where $M(<R)$  is the  total mass  within  a spherical
radius, $R$, the radius that corresponds to a given $j$ is given by
\begin{equation}
R = \frac{j^2}{G M(<j)}.
\end{equation}
This has a  number of desirable properties: 1) At  each time step it
is trivial  to calculate  how much  cold gas is added  to each  bin in
$j$. 2) Over time, as  the potential well changes, the specific angular
momentum of the baryons is automatically conserved; 3) The response of
the halo  to the  cooling of the  baryons is automatically  taken into
account.  4) The  resulting radial  grid is  adaptive, as  the mapping
between $j$ and $R$ depends on the amount of mass enclosed.

The  price that  we  pay for  these  advantages is  that the  circular
velocity due to the disk is  not calculated correctly. Due to the disk
geometry the  true circular velocity  deviates from that given  by the
spherically enclosed mass.   For example the peak $V_{\rm  circ}$ of a
thick exponential disk is $\simeq 10\%$ higher than that obtained using
the  enclosed mass,  and  at  small radii  the  proper $V_{\rm  circ}$
increases  linearly  with  radius,  whereas using  the  enclosed  mass
$V_{\rm  circ}$ scales as  $R^{1/2}$. 

However, given that computing the proper circular velocity of the disk
is very time  consuming, and that it is sensitive  to gradients in the
disk density profile, and that using it would remove the simplicity of
the $j$-grid  approach we feel  that it is  a price worth  paying. 

As we  show below  (in \S \ref{sec:vlr}),  and as discussed  in Dutton
\etal  (2007),  models with  halo  contraction  (and standard  stellar
IMF's) are unable to reproduce  the zero points of the $VMR$ relations
as well as  the low galaxy formation efficiency  required to reconcile
the halo mass function and  galaxy stellar mass function.  While there
are processes such as dynamical  friction and impulsive mass loss that
can expand the halo, implementing these in a galaxy evolution model is
a non-trivial task.   Thus for simplicity we wish  to consider a model
in which the dark halo does not respond to galaxy formation. 

Note that simply using the mapping between radius and $j$ based on the
halo profile  (at each time step)  in the absence  of galaxy formation
would not  conserve the specific  angular momentum of the  baryons. To
calculate the mapping between $j$ and  radius, for the case of {\it no
  adiabatic contraction}, we solve the equation
\begin{equation}
  R = \frac{j^2/G}{M_{\rm halo}(<R) + M_{\rm disk}(<j)},
\end{equation}
where $M_{\rm halo}(R)$ is the mass (within a spherical radius $R$) of
the dark matter plus hot gas  halo in the absence of galaxy formation,
and $M_{\rm  disk}(j)$ is the mass  of the disk (gas  plus stars) with
specific angular momentum less than $j$.  This way the self-gravity of
the disk is included but adiabatic contraction is ignored.

\begin{figure*}
\begin{center}
\centerline
{\psfig{figure=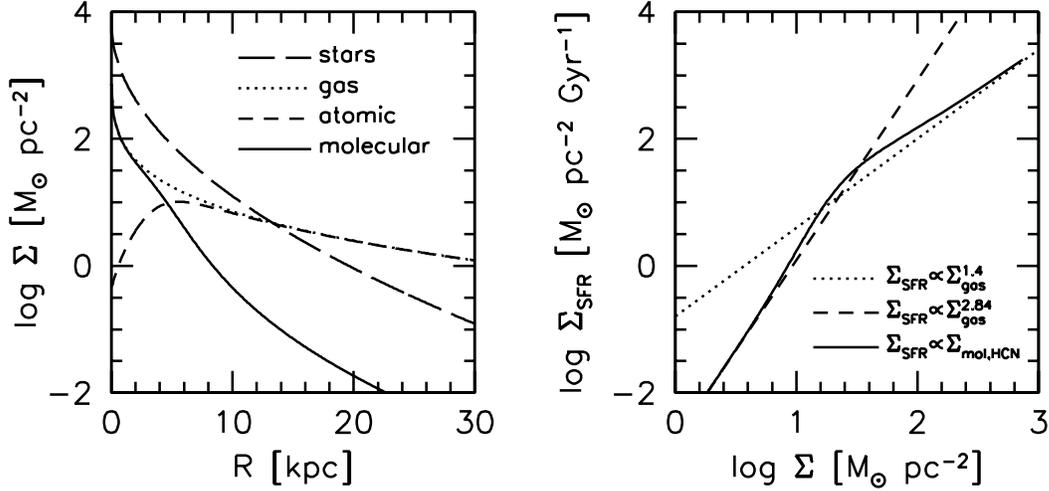,width=0.8\textwidth}}
\caption[Star  formation  laws]{\footnotesize  Right:  Star  formation
  surface density  vs total gas  surface density for our  adopted star
  formation  model   [Eq.~(\ref{eq:SKmol})],  with  $\tilde\epsilon_{\rm
    SF}=13\,\Gyr^{-1}$. This  has been calculated  using the molecular
  and total gas densities of the  model in the left panel.  The dotted
  line shows the standard SK relation, which our model converges to at
  high gas  density [Eq.~(\ref{eq:SKhigh})].  The dashed  line shows the
  relation   our    model   converges   to   at    low   gas   density
  Eq.~(\ref{eq:SKlow}). Left: Surface density profiles of stars and
  gas (total,  atomic and  molecular) which is  used to  calculate the
  right  panel.   The  molecular   fraction  is  computed  using  Eqs.~(\ref{eq:fmol}-\ref{eq:Pext}).}
\label{fig:SFR}
\end{center}
\end{figure*}

\subsection{Star Formation}

Observations have shown that the disk averaged star formation rates in
nearby spiral galaxies are well fitted by a simple Schmidt (1959) law
\begin{equation}
  \frac{\Sigma_{\rm SFR}}{[\Msun \,\rm pc^{-2}\, Gyr^{-1}]} = 
\frac{\epsilon_{\rm SF}}{[\Msun \, \rm pc^{-2}\, Gyr^{-1}]} \left(\frac{\Sigma_{\rm gas}}{[ 1 \Msun\,\rm pc^2]}\right)^n.
\end{equation}
Kennicutt  (1998) used  $\Sigma_{\rm gas}$  as the  total  gas density
(molecular   and  atomic,   but  not   including  helium)   and  found
$\epsilon_{\rm  SF}=0.25  \,\rm\Msun\,pc^{-2}\,Gyr^{-1}$, and  $n=1.4\pm0.15$
(Kennicutt  1998)\footnote{  Including  a  helium correction  of  1.36
  results   in  $\epsilon_{\rm   SF}=0.16\,   \rm  \Msun\,   pc^{-2}\,
  Gyr^{-1}$.}.  This  simple empirical law  holds over many  orders of
magnitude in  gas surface density,  and even applies  to circumnuclear
starburst regions.  However, when  applied to local gas densities, the
Schmidt law breaks  down at low gas densities,  corresponding to large
radii, where star formation has  been found to be abruptly suppressed.
Kennicutt  (1989) argued  that  this  suppression is  due  to the  gas
density falling  below the critical density for  global disk stability
as given by the Toomre criterion (Toomre 1964)
\begin{equation}
\Sigma_{\rm crit} = \sigma_{\rm gas} \kappa(R) / 3.36 G Q,
\end{equation}
where  $\sigma_{\rm  gas}$ is  the  velocity  dispersion  of the  gas,
$\kappa$  is  the  epicyclic  frequency,  and  $Q$  is  the  Toomre  Q
parameter.  Kennicutt (1989) found  that $\sigma_{\rm gas}=6 \kms$ and
$Q=1.5$  reproduces  the  observed  star formation  truncation  radii.
However other authors argue that  this is just a coincidence, and that
$\Sigma_{\rm crit}$ is not a threshold density (e.g. Schaye 2004).

The physical origin of the Schmidt-Kennicutt (SK) relation is also not
clear. However, something that is  well established is that stars form
out  of molecular  gas, and  predominantly in  giant  molecular clouds
(GMC's).  This  lead Wong \& Blitz  (2002) to argue for  a Schmidt law
based on the surface density of molecular gas.  For high gas densities
the molecular  gas dominates, so the two  prescriptions are identical.
However, for  low gas densities the molecular  fraction is suppressed,
resulting in a steep dependence  of the star formation rate density on
total  gas density.   This gives  an alternative  explanation  for the
suppression of star formation at low gas densities.

The fraction of  gas that is molecular, $f_{\rm  mol}$, can be defined
in terms  of the  mass ratio between  molecular and atomic  gas, $\cal
R_{\rm mol}$ by
\begin{equation}
\label{eq:fmol}
f_{\rm mol} = \frac{\cal R_{\rm mol}}{{\cal R}_{\rm mol}+1}.
\end{equation}
Empirically Wong \& Blitz (2002)  and Blitz \& Rosolowsky (2004; 2006)
have argued that ${\cal R}_{\rm mol}$ is determined to first order by the mid
plane pressure, $P_{\rm ext}$.  The most recent relation from Blitz \&
Rosolowsky (2006) is
\begin{equation}
\label{eq:Rmol}
{\cal R}_{\rm mol} = \frac{\Sigma_{\rm mol}}{\Sigma_{\rm atom}} 
= \left[ \frac{P_{\rm ext}/k}{4.3\pm0.6 \times 10^4} \right]^{0.92\pm0.1},
\end{equation}
where $k$ is Boltzmann's constant, and $P_{\rm ext}/k$ is in cgs units.
For a gas plus stellar disk the mid plane pressure is given, to
within 10\% by (Elmegreen 1993)
\begin{equation}
\label{eq:Pext}
P_{\rm ext} \simeq \frac{\pi}{2} G \Sigma_{\rm gas} \left [
  \Sigma_{\rm gas} + 
\left(\frac{\sigma_{\rm gas}}{\sigma_{\rm star}}\right) \Sigma_{\rm
  star} \right],
\end{equation}
where  $\sigma_{\rm gas}$  and  $\sigma_{\rm star}$  are the  velocity
dispersions of  the gas and stellar disk,  respectively. For simplicity
we  will assume  $\sigma_{\rm gas}/\sigma_{\rm  star}=0.1$ which  is a
reasonable  assumption  for  the   stellar  dominated  parts  of  disk
galaxies.  In the outer parts of disk galaxies this ratio is likely to
be  higher.  However, in  these  regions  gas  dominates, and  so  the
contribution of  the stars  to the mid  plane pressure  is negligible,
regardless of  the ratio  between $\sigma_{\rm gas}$  and $\sigma_{\rm
  star}$.    In   the   inner   regions   of   galaxies   $\sigma_{\rm
  gas}/\sigma_{\rm star}$ is  likely to be smaller than  0.1, but here
the densities are high enough that $f_{\rm mol} \simeq 1$.

Following  Blitz \& Rosolowski  (2006) we  assume that  star formation
takes place  in dense  molecular gas, traced  by HCN, with  a constant
star formation efficiency
\begin{equation}
\label{eq:SKmol}
\frac{\Sigma_{\rm SFR}}{[\Msun \rm  pc^{-2} Gyr ^{-1}]} = \frac{\tilde\epsilon_{\rm SF}}{[\rm Gyr^{-1}]} \frac{\Sigma_{\rm mol, HCN}}{[\Msun \rm pc^{-2}]},
\end{equation}
where $\tilde\epsilon_{\rm SF} \simeq  10-13 \rm {Gyr}^{-1}$ (Gao \& Solomon
2004, Wu \etal 2005).  Expressing  this equation in terms of the total
gas content:
\begin{equation}
\frac{\Sigma_{\rm SFR}}{[\Msun \rm  pc^{-2} Gyr ^{-1}]} = \frac{\tilde\epsilon_{\rm SF}}{[\rm Gyr^{-1}]} \frac{\Sigma_{\rm gas}}{[\Msun \rm pc^{-2}]} \,f_{\rm mol}\, {\cal R}_{\rm HCN},
\end{equation}
where ${\cal R}_{\rm HCN} = \Sigma_{\rm mol, HCN}/\Sigma_{\rm mol}$ is
the ratio between  the dense molecular gas (as traced  by HCN) and the
total molecular gas.

Based on the arguments and references in Blitz \& Rosolowski (2006) we
adopt the following relation for ${\cal R}_{\rm HCN}$
\begin{equation}
{\cal R}_{\rm HCN} = 0.1 \left( 1 + \frac{\Sigma_{\rm mol}}{[200 \, \Msun \rm pc^{-2}]} \right)^{0.4}. 
\end{equation}
In the low pressure,  low molecular density regime, ${\cal R}_{\rm HCN}\simeq
0.1$, and thus Eq.~(\ref{eq:SKmol}) asymptotes to
\begin{equation}
\label{eq:SKlow}
\frac{\Sigma_{\rm SFR}}{[\Msun \rm pc^{-2} \rm Gyr^{-1}]} = 
\frac{\tilde\epsilon_{\rm SF}}{[\rm Gyr ^{-1}]}\frac{0.013}{[\Msun\,\rm pc^{-2}]}\, \left(\frac{\Sigma_{\rm gas}}{[\Msun \rm pc^{-2}]}\right)^{2.84}. 
\end{equation}
In the  high pressure, high  molecular density regime,  ${\cal R}_{\rm
  HCN}   \propto  \Sigma_{\rm  mol}^{0.4}$,   and  eq.(\ref{eq:SKmol})
asymptotes to the familiar SK relation
\begin{equation}
\label{eq:SKhigh}
\frac{\Sigma_{\rm SFR}}{[\Msun \rm pc^{-2} \rm Gyr^{-1}]}  
= \frac{\tilde \epsilon_{\rm SF}}{[\rm Gyr^{-1}]}\frac{0.012}{[\Msun\,\rm pc^{-2}]}\, \left(\frac{\Sigma_{\rm gas}}{[\Msun \rm pc^{-2}]}\right)^{1.4}.
\end{equation}
Furthermore, with $\tilde\epsilon_{\rm SF}=13 \,\rm Gyr^{-1}$, we recover the
coefficient of $\epsilon_{\rm SF}=0.16$ of the standard SK relation.

Fig.~\ref{fig:SFR}  shows  the relation  between  star formation  rate
density   and    gas   density   for   our    star   formation   model
Eq.~(\ref{eq:SKmol}).   Note  that in  order  to compute  $\Sigma_{\rm
  SFR}$ one needs to  know $\Sigma_{\rm star}$ and $\Sigma_{\rm gas}$.
For illustrative  purposes we  have chosen a  stellar and  gas density
profile representative  of bright nearby disk galaxies  (left panel of
Fig.~\ref{fig:SFR}).  At  small radii, and high  gas densities $f_{\rm
  mol} \simeq 1$,  and the star formation law  follows the standard SK
relation.  At  large radii the  molecular fraction is very  low, which
results in a steeper slope to the star formation law.

We implement the star formation recipe given by Eq.(\ref{eq:SKmol}) as
follows.  At each time step and  annulus in the disk, we calculate the
star formation  rate.  Then we use the  following approximation (valid
for times  steps small compared to  the star formation  time scale) to
calculate the mass of newly formed stars
\begin{equation}
\Delta M_{\rm star}(R) = A(R) \,\Sigma_{\rm SFR}(R,t)\,\Delta t,
\end{equation}
with  $A$ the  area  of the  annulus,  and $\Delta  t$  the time  step
interval.

\subsection{Supernova Feedback}
When stars evolve they put  energy back into the inter stellar medium.
The effect of this on the  star formation rate is partially taken into
account by our empirically  determined star formation recipe.  What is
not taken  into account is a  feedback driven outflow of  gas from the
disk.  The  physical mechanism responsible  for driving outflows  is a
subject of debate (e.g. Finlator \& Dav\'e 2008), so in this paper we
consider both energy and momentum driven winds.

Following van den Bosch (2001) we assume that the outflow moves at the
local  escape  velocity  of  the  disk-halo system.   This  choice  is
motivated by  the fact that  it maximizes the  mass loss from  the 
disk-halo system (lower  velocity winds will not escape  the halo, and
higher  velocity winds will  carry less  mass).

For our  energy driven  wind model following  Dekel \& Silk  (1986) we
assume that  the kinetic energy  of the wind  is equal to  a fraction,
$\epsilon_{\rm EFB}$,  of the kinetic energy produced  by SN. However,
contrary to Dekel  \& Silk (1986) we apply  this energy condition {\it
  locally} in the  disk as a function of  radius, rather than globally
to the whole galaxy.  Thus the mass ejected from radius, $R$, during a
given time step is given by
\begin{equation}
\Delta M_{\rm eject}(R) = \frac{2\,\epsilon_{\rm EFB}\,E_{\rm SN}\,\eta_{\rm SN}}
{V_{\rm esc}^2(R)}\Delta M_{\rm star}(R).
\end{equation}
Here $\Delta M_{\rm  star}(R)$ is the mass in  stars formed at radius,
$R$,   $E_{\rm SN}=10^{51}$    erg   $\simeq   5.0\times    10^{7}   \,\rm
km^2\,s^{-2}\,  \Msun$   is  the  energy  produced  by   one  SN,  and
$\eta_{\rm SN}=8.3\times10^{-3}$  is the number  of SN  per solar  mass of
stars formed (for a Chabrier IMF).  

The local escape velocity is given
by
\begin{equation}
  V_{\rm esc} (R) = \sqrt{2 |\Phi_{\rm tot}(R)|},
\end{equation}
where $\Phi_{\rm  tot}$ is the sum  of the potentials due  to the disk
(stars plus gas) and halo (dark  matter plus hot gas), and is computed
assuming spherical symmetry.

For our momentum driven wind model  we assume that the momentum of the
wind is  equal to  a fraction, $\epsilon_{\rm  MFB}$, of  the momentum
produced by SN, thus the mass ejected from radius, $R$, during a given
time step is given by
\begin{equation}
  \Delta M_{\rm eject}(R) = \frac{\epsilon_{\rm MFB}\,p_{\rm SN}\,\eta_{\rm SN}}{V_{\rm esc}(R)}\Delta M_{\rm star}(R).
\end{equation}
Here $p_{\rm SN}=3\times 10^{4}\Msun\kms$  is the momentum produced by
one SN, assuming  that each SN produces $\simeq  10 \Msun$ of material
moving  at $v\simeq 3000  \kms$ (Murray,  Quataert \&  Thompson 2005).
Note  that this  corresponds to  a kinetic  energy of  $4.5\times 10^7
\,\rm \Msun \, km^{2}\,s^{-2}\simeq 10^{51}$ erg.

We assume that  the ejected mass is lost forever  from the system: the
ejected mass is not considered for later infall, and the corresponding
metals are  not used to enrich  the infalling gas.  This  is clearly a
dramatic oversimplification,  but we make this choice  to maximize the
amount of gas that is lost from the system.

\subsection{Stellar Populations and Chemical Enrichment}
In order to convert stellar  mass into luminosities we use the Bruzual
\& Charlot  (2003) stellar population synthesis  models.  These models
provide the luminosities $L(t,Z)$ of a single burst stellar population
with  a  total  mass of  $1\Msun$  as  a  function  of age,  $t$,  and
metallicity,  $Z$,  in  various  optical pass-bands.  To  compute  the
luminosities  of our model  stellar populations  we convolve  the star
formation  history  of our  galaxies  with  the  single burst  stellar
population synthesis models.

In order  to model the  chemical enrichment of  the ISM, we  adopt the
instantaneous  recycling  approximation.  We  assume  that a  fraction
$\cal{R}$ of the  mass in stars formed is  instantaneously returned to
the cold gas  phase with a yield $y$ (defined as  the fraction of mass
converted into stars that is returned  to the ISM in the form of newly
processed metals).

The equations for the change in the cold gas mass and metals are:
\begin{eqnarray}
\Delta M_{\rm cold} = \Delta M_{\rm  cool}
- (1-{\cal R})\Delta M_{\rm star} - \Delta M_{\rm eject}
\end{eqnarray}
\begin{eqnarray}
\Delta M_{\rm metal} = \Zhot\, \Delta M_{\rm cool} 
-Z_{\rm cold}(1-{\cal R}) \Delta M_{\rm star} \nonumber  
\\ + y \Delta M_{\rm star} -Z_{\rm cold} \Delta M_{\rm eject}.
\end{eqnarray}

The metallicity of the cold gas is then given by $Z_{\rm cold} = M_{\rm
  metal}/M_{\rm cold}$. Note we assume  that the ejected metals do not
enrich the hot halo gas.

\subsection{Book Keeping}
We now  briefly describe  how we  keep track of  the evolution  of the
various mass  components.  Given a  $z=0$ halo mass (dark  matter plus
hot  gas) and  concentration  we compute  the  MAH of  the halo  using
Eq.~(\ref{eq:MAH}).  The  evolution of the virial  radius and internal
structure  of the halo  is then  determined by  Eqs.~(\ref{eq:RMz}) \&
(\ref{eq:cz}).  We set  up a grid in radius from 0.001  to 1 times the
redshift zero virial radius.  As described in \S~\ref{sec:am}, for the
purposes of conserving angular momentum it is more convenient to use a
grid in specific  angular momentum, $j$.  Thus we  convert the grid in
$R$, to a grid in $j$ using $j^2/G = R M(R)$.

For each time step ($t-\Delta t,  t$) we compute the halo mass that is
added  to each  radial bin  so  that the  total mass  follows the  NFW
profile for a halo of a given $c(z)$ and $\Mvir(z)$.  Thus
\begin{equation}
\Delta M_{\rm vir}(j,t) = M_{\rm vir}(j,t) - M_{\rm vir}(j,t-\Delta
t). 
\end{equation}
We assume  that the baryons make  up a fraction $f_{\rm  bar}$ of this
mass, so that 
\begin{equation}
\Delta M_{\rm DM}(j,t) = (1-\fbar)\Delta M_{\rm vir}(j,t),
\end{equation}
\begin{equation}
\Delta M_{\rm hot}(j,t) = \fbar\Delta M_{\rm vir}(j,t).
\end{equation}
When  we compute  the circular  velocity we  assume that  the  hot gas
follows the mass distribution of  the dark matter.  When computing the
cooling time we  assume $\rho_{\rm hot}$ is $\fbar$  times the density
at the virial radius at the time when the gas virialized.

At each time  step we then compute, using the  recipes in the previous
sections, the  amount of gas  that cools, $\Delta  M_{\rm cool}(j,t)$,
the amount of stars formed  $\Delta M_{\rm star}(j,t)$, and the amount
of  ejected   gas  $\Delta  M_{\rm  eject}(j,t)$.    For  the  stellar
population modeling we keep track of  the mass of stars formed at each
time step and the metallicity of  the gas from which the stars formed,
$Z_{\rm cold}(j,t)$.

\subsection{Overview of Model Parameters}
\label{sec:param}
The input parameters of our models are as follows.

(1) Cosmology:  $\Omega_{\rm m,0}, \OmegaL,  \Omega_{\rm b}, \sigma_8,
h, n$.  In this paper we adopt a flat $\Lambda$CDM cosmology motivated
by the 5th  year WMAP results (Dunkley \etal  2009), with $\Omega_{\rm
  m,0}=0.258,    \OmegaL=0.742,     \Omega_{\rm    b}=0.044,    h=0.7,
\sigma_8=0.80$ and $n=1$.
    
(2)  Halo structure:  $K, F,  \sigma_{\ln c}$.   We adopt  the Bullock
\etal (2001a) model with $F=0.01$, $K=3.7$, and $\sigma_{\ln c}=0.25$.
These parameters reproduce the  distribution of halo concentrations of
relaxed dark  matter haloes  in numerical simulations  (Wechsler \etal
2002; Macci\`o \etal 2007, 2008).

(3)  Angular   momentum  distribution:  $\bar{\lambda}$,  $\sigma_{\ln
  \lambda}$, $\alpha$,  $\sigma_{\log\alpha}$.  As fiducial  values we
adopt  a median  spin parameter  $\bar{\lambda}=0.035$ with  a scatter
$\sigma_{\ln \lambda}=0.54$,  corresponding to relaxed  haloes (Macci\`o
\etal 2007, 2008).  For the  angular momentum shape parameter we adopt
a  median $\bar{\alpha}=0.90$, and  scatter $\sigma_{\log\alpha}=0.11$
(Sharma \& Steinmetz 2005).

(4) Star formation: $\tilde\epsilon_{\rm SF}$. We use a star formation
model based  on dense molecular gas  [Eq.~(\ref{eq:SKmol})], and adopt
$\tilde\epsilon_{\rm SF}=13\, \rm Gyr^{-1}$.

(5) Feedback:  $\epsilon_{\rm EFB}$, $\epsilon_{\rm  MFB}$, $\eta_{\rm
  SN}$,  $E_{\rm SN}$,  $p_{\rm SN}$.   We adopt  $E_{\rm SN}=10^{51}$
erg,  $p_{\rm  SN}=3\times  10^{4}   \,  \Msun  \,  \kms$,  $\eta_{\rm
  SN}=8.3\times10^{-3}$.    We    treat   $\epsilon_{\rm   EFB}$   and
$\epsilon_{\rm MFB}$ as free parameters.
    
(6) Stellar populations and  chemical enrichment: ${\cal R}, y, Z_{\rm
  hot}$, and the choice of  initial mass function (IMF).  We adopt the
Chabrier (2003) IMF, a  return  fraction ${\cal  R}=0.35$,  a stellar  yield
$y=0.02$,  and a  metallicity of  the hot  gas of  $Z_{\rm hot}=0.002$
$(\simeq 0.1 Z_{\odot})$.

\subsection{Overview of  Output Parameters}
The output  parameters of our models,  that we discuss  in this paper,
are as follows:

\begin{itemize}

\item $\Mvir$, total mass inside virial radius [$\Msun$]
\item $\Vvir$, circular velocity at the virial radius [$\kms$]
\item $\Vmaxh$, maximum circular velocity of the halo without galaxy formation [$\kms$]
\item $\Mgal$, galaxy mass (stars and cold gas) [$\Msun$]
\item $\Mstar$, stellar mass [$\Msun$].
\item $\Mcold$, cold gas mass [$\Msun$].
\item $Z_{\rm cold}$, metallicity of cold gas.
\item $\epsilon_{\rm GF}=\mgal/(\Omega_{\rm b}/\Omega_{\rm m,0})$, galaxy formation efficiency.
\item $\mgal=\Mgal/\Mvir$, galaxy mass fraction.
\item $\mstar=\Mstar/\Mvir$, stellar mass fraction.
\item $\mcold=\Mcold/\Mvir$, cold gas mass fraction.
\item $f_{\rm gas}=\Mcold/(\Mcold + \Mstar)$, cold gas fraction.
\item $V_{2.2I}$, the circular velocity measured at 2.15 $I$-band disk
  scale lengths [$\kms$].
\item $R_{\rm dI}$, $I$-band disk scale length [$\kpc$].
\item $\jgal=\Jgal/\Jvir$, galaxy angular momentum fraction.
\item $\jstar=\Jstar/\Jvir$, stellar angular momentum fraction. 
\item $\jcold=\Jcold/\Jvir$, cold gas  angular momentum fraction.
\item $\lamgal=\lambda (\jgal/\mgal)$, galaxy spin parameter. 
\item $\lamstar=\lambda (\jstar/\mstar)$, stellar spin parameter.
\item $\lamcold=\lambda (\jcold/\mcold)$, cold gas spin parameter.
\item $\dot{M}_{\rm bar}=f_{\rm bar}\dot{M}_{\rm vir}$, the baryonic mass accretion rate [$\Msun \,\rm Gyr^{-1}$]
\item $\dot{M}_{\rm cool}$, the mass cooling rate [$\Msun \,\rm Gyr^{-1}$]
\item $\dot{M}_{\rm SF}$, the star formation rate [$\Msun \,\rm Gyr^{-1}$]
\item $\dot{M}_{\rm w}$, the mass outflow rate [$\Msun \,\rm Gyr^{-1}$]
\item $\eta=\dot{M}_{\rm w}/\dot{M}_{\rm sf}$, the mass loading factor
\item   $V_{\rm  w}=\Sigma_i   V_{\rm  w}(i)   \dot{M}_{\rm   w}(i)  /
  \dot{M}_{\rm w}$, the (outflow)  mass weighted mean outflow velocity
  [$\kms$], where the $i$ refers to the radial grid position.
\end{itemize} 

The disk  scale lengths are determined using  the following procedure,
which was  developed to  give robust disk  scale lengths for  the full
range of surface brightness profiles  produced by our model.  We first
compute the local disk scale length between the radii enclosing 50 and
99\% of the  stellar mass. The local disk scale  length is computed at
radial bin $i$ by using the surface densities and radii at radial bins
$i-1$ and $i+1$. We then determine the maximum of the local disk scale
length,  and the  radius  where this  maximum  occurs, $R_{\rm  max}$.
Finally we determine  the scale length of the disk  using a linear fit
to the  model data over  the range  $0.6 R_{\rm max}  \le R \le 1.6
R_{\rm max}$.

\subsection{Comparison with other Disk Galaxy Structure Models}

In this section we place our model in the context of existing analytic
and  semi-analytic models  of disk  galaxy structure/formation  in the
literature.   We classify  these  models into  two  general types:  1)
models that  conserve {\it total} specific angular  momentum (i.e. the
structural profile of the disk is assumed) and 2) models that conserve
the  {\it  distribution}  of   specific  angular  momentum  (i.e.  the
structural profile of the disk  is derived).  Both of these classes of
models can  be static or include evolution.   The essential assumption
in  both  classes  of  models  is  that the  disk  is  in  centrifugal
equilibrium inside  some potential (which  may or may not  include the
self gravity of the disk).

\subsubsection{Models that Conserve Total Specific Angular Momentum}
In the simplest models of this class, the circular velocity is assumed
to be constant, i.e.   corresponding to an isothermal density profile,
and  the self  gravity  of the  disk  is ignored.   This  model has  3
parameters:  the circular velocity,  the spin  parameter and  the disk
mass fraction. Such a model was  discussed in Mo, Mao, \& White (1998;
MMW), and  is widely  used in Semi  Analytic Models  (e.g.  Kauffmann,
White,  \& Guiderdoni  1993; Cole  \etal 1994;  Somerville  \& Primack
1999; Hatton \etal 2003; Croton \etal 2006).

A more  realistic version of this  model includes the  self gravity of
the  disk and  adiabatic  contraction of  the  halo (Blumenthal  \etal
1986),  which  usually results  in  smaller  sizes  for a  given  spin
parameter and  disk mass  fraction.  In Mo,  Mao \& White  (1998), the
halo was assumed to  be an NFW profile, and the disk  was assumed to be
exponential.  This  model has 4  parameters: the circular  velocity of
the halo,  the concentration of the  halo, the spin  parameter and the
disk mass fraction.   This version of the MMW model  is widely used in
Semi  Analytic  Models  (Cole   \etal  (2000);  Benson  \etal  (2003);
Somerville \etal  2008) and studies  of disk galaxy  scaling relations
(e.g.   Navarro \& Steinmetz  2000; Pizagno  \etal 2005;  Dutton \etal
2007; Gnedin \etal 2007).

\subsubsection{Models that Conserve  the Distribution  of   Specific  Angular
  Momentum}
The  MMW type model  is useful  for understanding  the origin  of disk
galaxy scaling  relations, but it does  not explain the  origin of the
density  profiles of  galaxy disks,  or the  relation between  gas and
stars in galaxy disks. In  order to address these questions, one needs
to start from some  specific angular momentum distribution (AMD). This
AMD may be that of a sphere in solid body rotation, or preferably that
found in  cosmological simulations  (e.g.  Bullock \etal  2001b).  The
radial  density profile  of the  disk is  then determined  by detailed
conservation of specific angular momentum.

As with the MMW type models, these models may ignore the self gravity
of the disk by assuming the total density profile is isothermal
(e.g. Ferguson \& Clarke 2001), or include the self gravity of he disk
inside a dark matter halo (e.g.  Dalcanton, Spergel \& Summers 1997).
These models may also include evolution of the baryonic disk, by
following the cooling of gas inside growing dark mater haloes, and
evolution of the stellar disk, by following the star formation as a
function of radius (e.g.  Firmani \& Avila Reese 2000; van den Bosch
2001, 2002; Stringer \& Benson 2007).

Our model fall into this later category, being an evolution of the van
den Bosch (2001; 2002) models.   The strength of these models over the
MMW type  models is that they  can be used  to self-consistently study
the origin  and evolution of  disk density profiles (stars,  gas, star
formation  rates, inflows,  outflows, metallicity,  stellar  ages) and
rotation curve shapes.


\section{The dependence of the Galaxy  Velocity - Mass - Size relations on feedback}
\label{sec:vlr}
In this section we investigate the dependence of the scaling relations
between rotation velocity, stellar mass and stellar disk scale size on the
feedback model.

\begin{figure*}
\centerline{
\psfig{figure=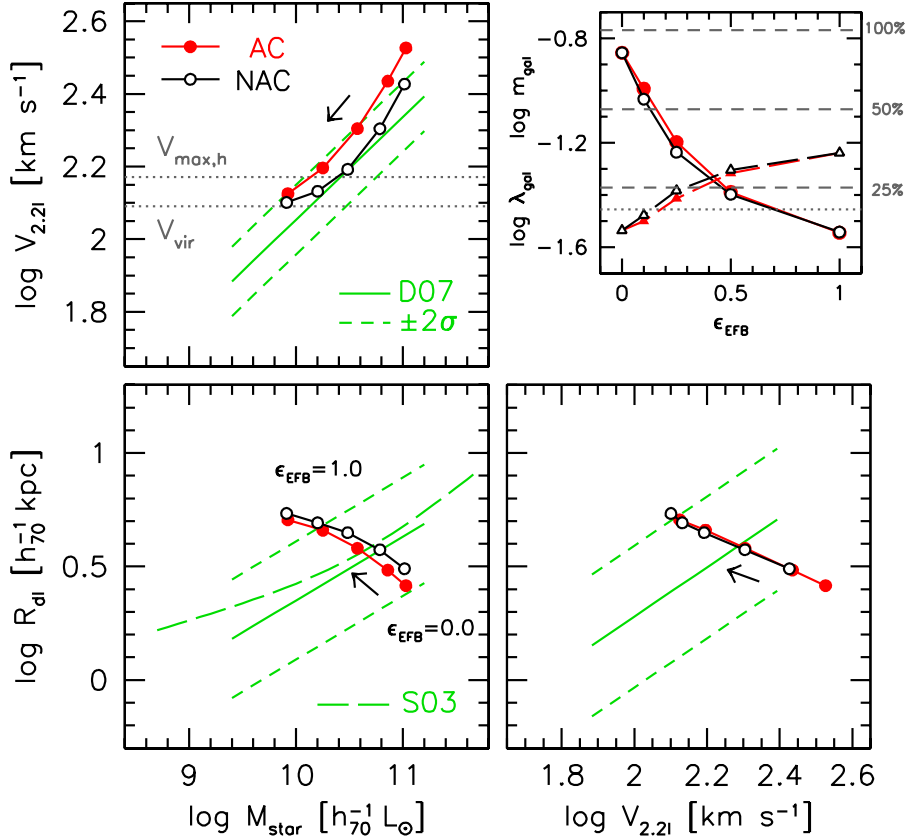,width=0.7\textwidth}}
\caption[]{\footnotesize Effect of energy feedback efficiency,
  $\epsilon_{\rm EFB}$, on the position of a galaxy with $\Mvir = 6.3
  \times 10^{11} \hMsun$ in the $VMR$ planes. Models have feedback
  efficiencies of $\epsilon_{\rm EFB}=0, 0.1, 0.25, 0.5$, and $1.0$.
  The arrows indicate the direction of increasing $\epsilon_{\rm
    EFB}$.  Models with adiabatic contraction are shown with solid red
  symbols, models without adiabatic contraction are shown as black
  open symbols.  In the $VM$ plane the horizontal dotted grey lines
  show virial velocity of the halo, $\Vvir$, and the maximum circular
  velocity of the halo prior to galaxy formation, $\Vmaxh$.  The solid
  and dashed green lines in the $VMR$ panels show the mean and
  2$\sigma$ scatter of the observed relations from Dutton \etal 2007
  (D07), assuming a Chabrier IMF. The long dashed green line shows the
  observed half-light radius stellar mass relation from Shen \etal
  (2003).  The panel in the top right shows the effect of feedback on
  the galaxy mass fraction, $\mgal$ (circles), and galaxy spin
  parameter, $\lamgal$ (triangles).  The dashed horizontal lines show
  galaxy formation efficiencies of 100, 50, and 25 percent, the dotted
  horizontal line shows the spin parameter of the halo. As the
  feedback efficiency is increased the galaxy mass fraction ($m_{\rm
    gal}$) decreases, the galaxy spin parameter ($\lambda_{\rm gal}$)
  increases, the rotation velocity decreases, the stellar mass
  decreases, and the size of the stellar disk increases.}
\label{fig:VMRIm-efb}
\end{figure*}

\begin{figure*}
\centerline
{\psfig{figure=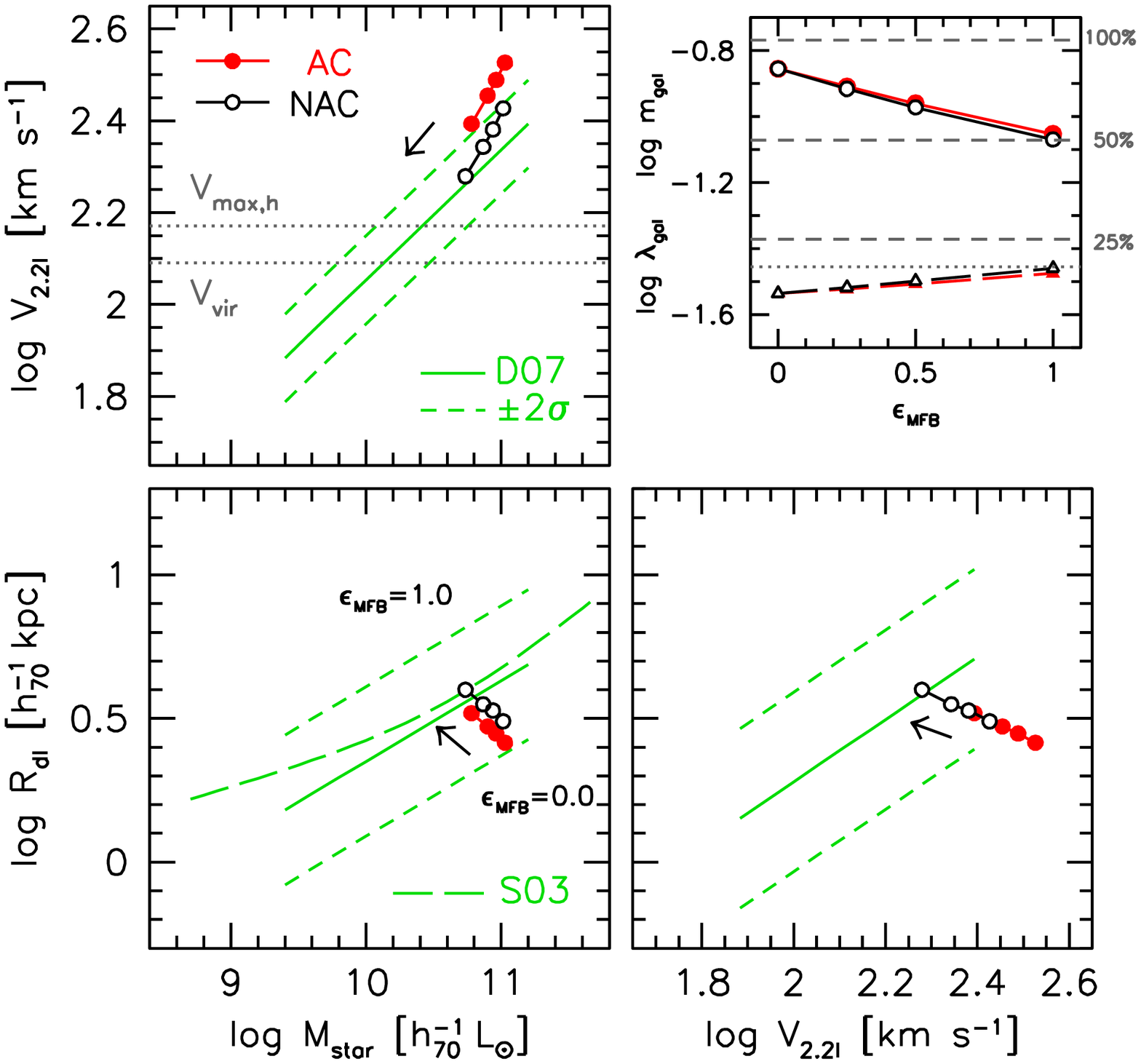,width=0.7\textwidth}}
\caption[]{\footnotesize
    As fig \ref{fig:VMRIm-efb} but for momentum driven winds.}
\label{fig:VMRIm-mfb}
\end{figure*}

\subsection{Observed Disk Galaxy Scaling Relations}
\label{sec:obsvlr}
Here we overview the main observed velocity, mass, size, scaling
relations that we are going to compare our models to.  We use the
relations between rotation velocity at 2.2 $I$-band disk scale
lengths, $V_{2.2I}$, stellar mass, $M_{\rm star}$, and $I$-band disk
scale length, $R_{{\rm d}I}$, from the data set of Courteau \etal
(2007), as presented in Dutton \etal (2007).  The stellar masses in
Dutton \etal (2007) were derived from $I$-band luminosities using the
relations from Bell \etal (2003) corresponding to a diet Salpeter
IMF. Here we adopt a Chabrier IMF, and so subtract 0.10 dex from the
stellar masses.

The stellar mass TF relation is given by
\begin{equation}
\log \frac{V_{2.2I}}{[\kms]} = 2.195 + 0.259\, \left(\log \frac{\Mstar}{[h_{70}^{-2} M_{\odot}]} -10.5\right),
\end{equation}
the size-stellar mass relation is given by
\begin{equation}
\log \frac{R_{{\rm d}I}}{[h_{70}^{-1} \rm kpc]} 
= 0.491 + 0.281\,\left(\log \frac{\Mstar}{[h_{70}^{-2} M_{\odot}]} -10.5\right),
\end{equation}
and the corresponding size-velocity relation is given by
\begin{equation}
\log \frac{R_{{\rm d}I}}{[h_{70}^{-1} \rm kpc]} 
= 0.491 + 1.086\,\left(\log \frac{V_{2.2I}}{[\kms]} -2.195\right).
\end{equation}
The  intrinsic   scatter  in  these  relations  is   estimated  to  be
$\sigma_{\log_{10}   V|M}\simeq0.05$,   $\sigma_{\log_{10} R|M}\simeq0.13$,   and
$\sigma_{\log_{10} R|V}\simeq0.16$. 
  The errors on the slopes  of the $VM$,
$RM$, and  $RV$ relations from  fitting uncertainties are  0.01, 0.02,
0.12,    respectively.    However,   systematic    uncertainties   are
significantly  larger, and  harder to  quantify. The  most significant
selection  effect  for the  slope  of  the  $RM$ relation  is  surface
brightness.  The data set compiled  by Courteau \etal (2007) is likely
missing lower surface brightness galaxies, and thus over-estimates the
slope of  the size-mass relation at  low masses. Such  a conclusion is
supported   by  Shen  \etal   (2003)  who   studied  the   half  light
radius-stellar  mass  relation  ($R_{50}-\Mstar$)  for a  much  larger
sample of galaxies ($\sim 10^5$) from the SDSS.  They find a log slope
of 0.14 at  low masses, increasing to 0.39 at  high masses.  

\subsection{A Fiducial Model}
To illustrate the effect that feedback has on observable properties of
disk galaxies  we consider a  model with virial  mass $\Mvir=6.3\times
10^{11}\hMsun$,   and  median   concentration  and   angular  momentum
parameters: $\cvir=9.9$, $\lambda=0.035$ and $\alpha=0.9$. The results
of varying the energy and momentum feedback efficiency parameters from
0 to 1 are shown in Figs.  \ref{fig:VMRIm-efb} \& \ref{fig:VMRIm-mfb},
respectively.  The main panels show the $VMR$ relations with the solid
line showing the mean relations from observations and the dashed lines
showing  the  2$\sigma$ scatter.   The  upper  right  panels in  these
figures shows  the more theoretical parameters  $\mgal$ and $\lamgal$.
These are  not directly observable  because they require  knowledge of
the  halo  mass, a  quantity  that  cannot,  at present,  be  reliably
measured for individual galaxies.

We  first focus  on  the  models with  energy  feedback and  adiabatic
contraction (solid red  points and lines in Fig.~\ref{fig:VMRIm-efb}).
A model with no feedback  ($\epsilon_{\rm EFB}=0$) results in a galaxy
mass fraction  of $\simeq 0.8\fbar$  (where $\fbar\simeq 0.17$  is the
universal baryon  fraction). The  galaxy formation efficiency  is less
than 1  because cooling  starts to become  inefficient at  late times.
Since the highest specific angular momentum material is accreted last,
and  this does  not  have time  to cool,  the  spin of  the galaxy  is
slightly  lower than  that of  the  halo.  The  high galaxy  formation
efficiency results  in a disk scale  length a factor of  $\simeq 1.8 $
too small and  a circular velocity at 2.2 disk  scale lengths a factor
of $\simeq 1.5$ too high.

When feedback is  included, some of the cold  baryons are ejected from
the disc (and  halo).  This trivially results in  lower $\mgal$ (hence
lower baryonic  mass) but  also, non-trivially, higher  $\lamgal$. The
higher $\lamgal$  is due to  the preferential ejection of  low angular
momentum   material,   which   we    discuss   in   more   detail   in
\S\ref{sec:mgal}.   Both of  these  effects results  in larger,  lower
surface density disks, which  result in less efficient star formation,
and  hence  higher  gas  fractions  and  lower  stellar  masses.   The
reduction in the amount and density  of the baryons results in a lower
rotation  velocity, both because  the baryons  contribute less  to the
circular velocity at 2.2 disk  scale lengths and because there is less
halo  contraction.    With  a  high  energy   feedback  efficiency  of
$\epsilon_{\rm  EFB}=1.0$  the galaxy  formation  efficiency drops  to
$\simeq 15\%$:  $\simeq 20\%$  of the baryons  have not  cooled, while
$\simeq 65\%$ of the baryons have been ejected from the disk and halo.
This model also has a galaxy spin a factor of $\simeq 1.7$ higher than
the halo  spin. The  low galaxy mass  fraction and higher  galaxy spin
parameter results in sizes that are  more than a factor of $\simeq 2 $
too  large. The feedback  efficiency can  be tuned  so that  the model
reproduces the size-mass, or  size-velocity relation. However, for all
feedback efficiencies, the  models rotate too fast at  a given stellar
mass.   This is because  higher feedback  efficiency results  in lower
stellar  masses as  well as  lower rotation  velocities, with  the net
result that galaxies move almost parallel to the TF relation.

\begin{figure}
\begin{center}
\psfig{figure=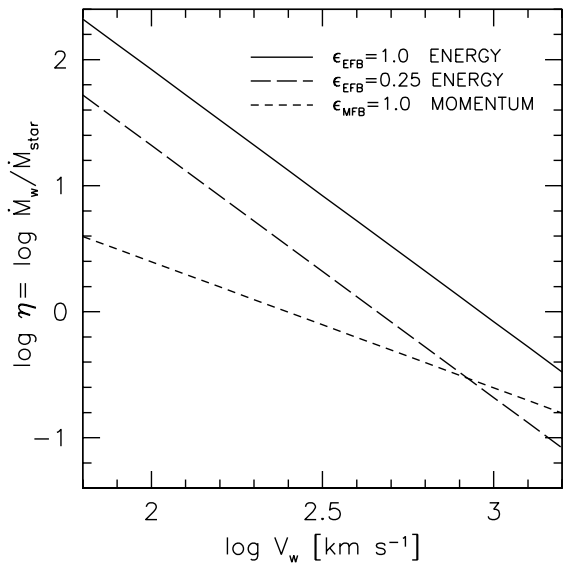,width=0.45\textwidth}
\caption[]{\footnotesize Mass loading  factor versus wind velocity for
  energy and  momentum driven winds. For equal  efficiencies an energy
  driven wind  is always more  efficient than a momentum  driven wind,
  especially for low wind velocities.}
\label{fig:etav}
\end{center}
\end{figure}

Fig.~\ref{fig:VMRIm-mfb}     shows     the     same     results     as
Fig.~\ref{fig:VMRIm-efb},  but  for the  momentum  driven wind  model.
Even with 100 per cent  efficiency this galaxy formation efficiency is
$\simeq 50\%$,  where $\simeq 30\%$  of the baryons have  been ejected
from the disk  and halo.  This results in sizes  that are in agreement
with observations, but  the models still rotate too  fast.  The reason
that  energy driven  winds are  more efficient  at ejecting  mass than
momentum driven winds is shown in Fig.~\ref{fig:etav}, which shows the
mass  loading  factor,  $\eta$,  versus  the  wind  velocity,  $V_{\rm
  w}$. The  mass loading  factor is defined  as the ratio  between the
mass  outflow rate  and the  star formation  rate.  For  energy driven
winds $\eta \propto V_{\rm w}^{-2}$, whereas for momentum driven winds
$\eta \propto  V_{\rm w}^{-1}$.   Thus everything being  equal, energy
driven  winds have higher  mass loading  factors than  momentum driven
winds for all typical wind velocities

\subsection{The Tully-Fisher Zero Point Problem}
A  common problem  to  both feedback  models,  for all  values of  the
feedback  efficiency,   is  that  they  over   predict  the  rotation
velocities.  This is a standard problem for galaxy formation models in
$\Lambda$CDM.  As  discussed in Dutton  \etal (2007) and  Gnedin \etal
(2007),  there are  3 solutions:  1) Lower  the  stellar mass-to-light
ratio (i.e.  for a given luminosity  there is less  stellar mass which
shifts observed galaxies to the  left and hence higher velocity in the
$VM$  plane)\footnote{Note  that  the  stellar  masses  of  the  model
  galaxies are  not completely independent  of the IMF, as  the return
  fraction is  IMF dependent. However, changes in  the return fraction
  are compensated  for by changes in  the star formation  rate. }.  2)
Lower   the  initial   halo  concentration   (which   directly  lowers
$V_{2.2}$).   3)  Reverse  halo  contraction  (which  directly  lowers
$V_{2.2}$).

A small change  in the stellar mass-to-light ratio  would be plausible
due to  systematic uncertainties  (such as in  the IMF or  the stellar
populations   synthesis  models)   in  the   measurement   of  stellar
mass-to-light ratios. However,  the stellar mass-to-light ratios would
have to be lowered by about a  factor of 2 to match the TF zero point.
Such a  large change would  require a top-heavy  IMF.  But all  of the
available constraints  on stellar mass-to-light ratios  point to IMF's
similar to Chabrier (e.g. de Jong \& Bell 2007).

Lower halo  concentrations would require  less power on  galaxy scales
than in standard  $\Lambda$CDM.  This would also reduce  the amount of
substructure,  which could  help solve  the missing  satellite problem
(Klypin \etal  1999; Moore \etal 1999). However,  the recent discovery
of  many satellite  galaxies around  the  Milky Way  has lessened  the
discrepancy  between  observations  and  standard  $\Lambda$CDM  (e.g.
Tollerud  \etal  2008  and  references  therein).   Furthermore  using
cosmological simulations with parameters  from the latest WMAP results
(Spergel  \etal 2007; Dunkley  \etal 2009),  Macci\`o \etal  (2008) have
shown that the central densities  of dark matter haloes are consistent
(both in normalization and scatter) with those measured from dwarf and
LSB  galaxies (which  typically  have maximum  rotation velocities  of
$\simeq 100 \kms$).  Thus there  does not seem compelling evidence for
a modification to $\Lambda$CDM on small scales.

Given that  reducing stellar mass-to-light ratios or  the initial halo
concentrations  do   not  seem   plausible,  we  consider   the  third
possibility, that  halos do not  contract as expected.  There  are two
processes that could cause the  halo to expand.  1) Dynamical friction
between baryons and the halo, e.g.  by infalling baryonic clumps (e.g.
El Zant \etal  2001; El-Zant \etal 2004; Tonini  \etal 2006; Jardel \&
Sellwood  2009);   or  by  galactic  bars  (Weinberg   \&  Katz  2002;
Holley-Bockelmann,  Weinberg \&  Katz 2005;  Sellwood 2008);  2) Rapid
(i.e.  non-adiabatic)  mass loss  from the galaxy,  e.g. by  SN driven
winds  (Gnedin \& Zhao  2002; Read  \& Gilmore  2005).  Both  of these
effects have been shown to be effective at expanding the halo, but the
combined effect (which  may be greater than the sum  of its parts) has
so  far not  been investigated.   Note  that our  standard model  with
adiabatic contraction takes into account the adiabatic expansion of the
halo due to outflows. However, since  there is a net inflow of baryons
into the centers of galaxies, the overall effect is halo contraction.

Furthermore halo contraction  is based on the idea  that galaxies form
by cooling  flows.  The hot gas  radiates away its  binding energy, so
when it  falls to the center of  the potential to form  the galaxy the
halo has to contract.  However, recent simulations have indicated that
the gas, in haloes that host  disk galaxies, is accreted by cold flows
(i.e.  it does not shock heat when it enters the virial radius).  This
new scenario  thus allows the possibility of  the baryons transferring
energy to the halo during galaxy formation.

The black circles  in Figs. \ref{fig:VMRIm-efb} \& \ref{fig:VMRIm-mfb}
show the  results for  two models without  halo contraction.  For both
energy and  momentum driven winds  the galaxy mass fractions  and spin
parameters  have the  same dependence  on the  feedback  efficiency in
models with and without halo contraction.  However, the models without
halo  contraction   have  significantly  lower   rotation  velocities,
allowing a match to the TF zero point for energy feedback efficiencies
of  $\simeq 0.1-0.5$,  or  momentum feedback  efficiencies of  $\simeq
1$. A  model with $\epsilon_{\rm EFB}=0.25$ also  has $V_{2.2I} \simeq
\Vmaxh$, and a galaxy formation efficiency of $\simeq 35\%$, consistent
with observations (see \S 1).   However, this model has sizes that are
too large. The solution to this  is to lower the spin parameter of the
baryons.   This could  occur if  disk galaxies  formed in  haloes with
lower than average  halo spin, or by the  baryons transferring some of
their angular momentum to the halo during galaxy assembly, for example
via dynamical friction.  As we  discussed above, such a process may be
responsible for expanding  the halo. As we show below,  a model with a
factor 1.4 lower halo spin  reproduces both $VM$ and $RM$ relations as
well  as low  galaxy  formation efficiencies.   The momentum  feedback
model with  maximum feedback efficiency,  $\epsilon_{\rm MFB}=1.0$, on
the other hand, roughly reproduces  the $VM$ and $RM$ zero points, but
it has galaxy  mass fractions and $V_{2.2I}/\Vmaxh$ that  are too high
compared to observational and theoretical constraints.

\begin{table}
 \begin{minipage}{0.4\textwidth}
  \caption{Model Parameters}
  \begin{tabular}{lccccc}
\hline  Name &  $\epsilon_{\rm EFB}$ & $\epsilon_{\rm MFB}$ & $\bar{\lambda}$& $\sigma_{\ln \lambda}$ & AC \\
I: No Feedback: & 0.0 & 0.0 & 0.035 & 0.35 & N \\ 
II: Momentum:& 0.0 & 1.0 & 0.035 & 0.35 & N \\ 
III: Energy: & 0.25 & 0.0 & 0.025 & 0.35 & N \\ 
\hline 
\label{tab:params}
\end{tabular}
\end{minipage}
\end{table}

\begin{figure*}
\centerline{
\psfig{figure=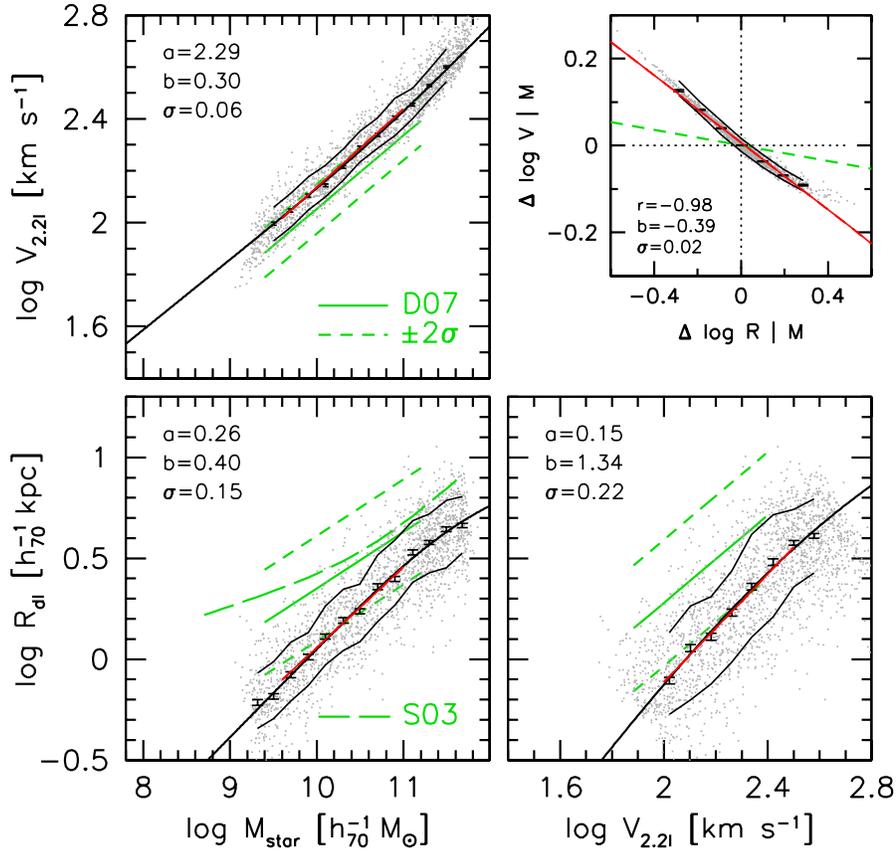,width=0.7\textwidth}}
\caption[]{\footnotesize $V_{2.2\rm I}-\Mstar-R_{\rm dI}$ relations
  for model I (no feedback) for virial masses logarithmically sampled
  in the range $10^{10} < \Mvir < 10^{13} \hMsun$.  The grey points
  and black lines show the models while the green lines show the
  observed relations. In the $VMR$ panels the solid and dashed green
  lines show the mean and 2$\sigma$ scatter, assuming a Chabrier IMF,
  of the observed relations from Dutton \etal 2007 (D07).  In the $RM$
  panel the long dashed green line shows the observed half-light
  radius stellar mass relation from Shen \etal (2003); In the $VM-RM$
  residual panel the green dashed line shows the observed correlation
  from Avila-Reese \etal (2008). The black lines show the 16th, and
  84th percentiles of the models, while the error bars show the
  Poisson error on the median. The thick black line shows a double
  power law fit to the median using the function in
  Eq.(\ref{eq:power1}), the parameters of the fits are given in
  Table~\ref{tab:power1}.  The solid red lines show the mean of the
  models fitted over a range corresponding to $10^{9.6} < \Mstar <
  10^{11.0} \,h_{70}^{-1}\Msun$.  The parameters of the fits (zero
  point, slope, scatter (in $\log_{10}$ units)) are given in the top
  left corner of each panel.  The relations are fitted as follows:
  $\log V=a+b\,(\log M -10.5)$; $\log R =a+b\,(\log M -10.5)$; $\log
  R=a+b\,(\log V -2.195)$.  The upper right panel shows the residuals of
  the $VM$ relation vs the residuals of the $RM$ relation.  The red
  lines show the mean and 1$\sigma$ scatter of a fit of the form:
  $\Delta \log V |M = b\, \Delta \log R |M$.  The correlation
  coefficient, $r$, is also given where $r=b \,\sigma_x / \sigma_y$,
  where $x=\Delta\log R | M$ and $y=\Delta \log V | M$. This model
  fails to reproduce all of the observations (with the exception of
  the slope of the $VM$ relation).}
\label{fig:VMRId-nfb-nac}
\end{figure*} 

\begin{figure*}
\centerline{
\psfig{figure=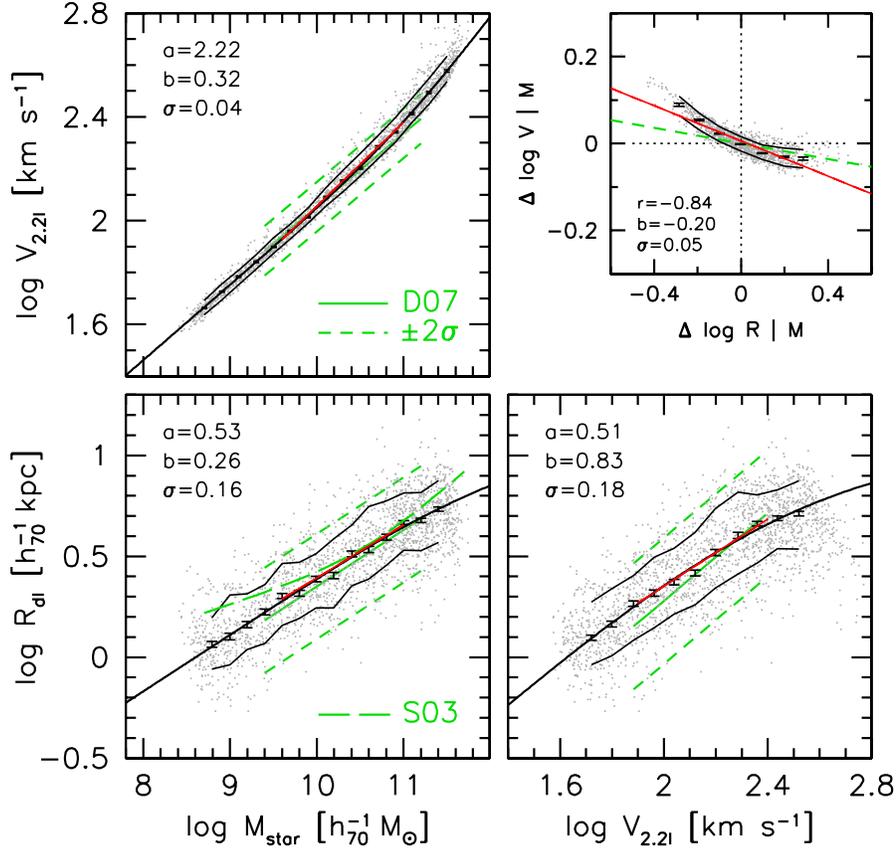,width=0.7\textwidth}}
\caption[]{\footnotesize As  Fig.~\ref{fig:VMRId-nfb-nac} but for Model
  II (momentum driven feedback). This model provides a reasonable match to the zero points, slopes and scatters of the $VMR$ relations.}
\label{fig:VMRId-mfb-nac}
\end{figure*} 

\begin{figure*}
\centerline{
\psfig{figure=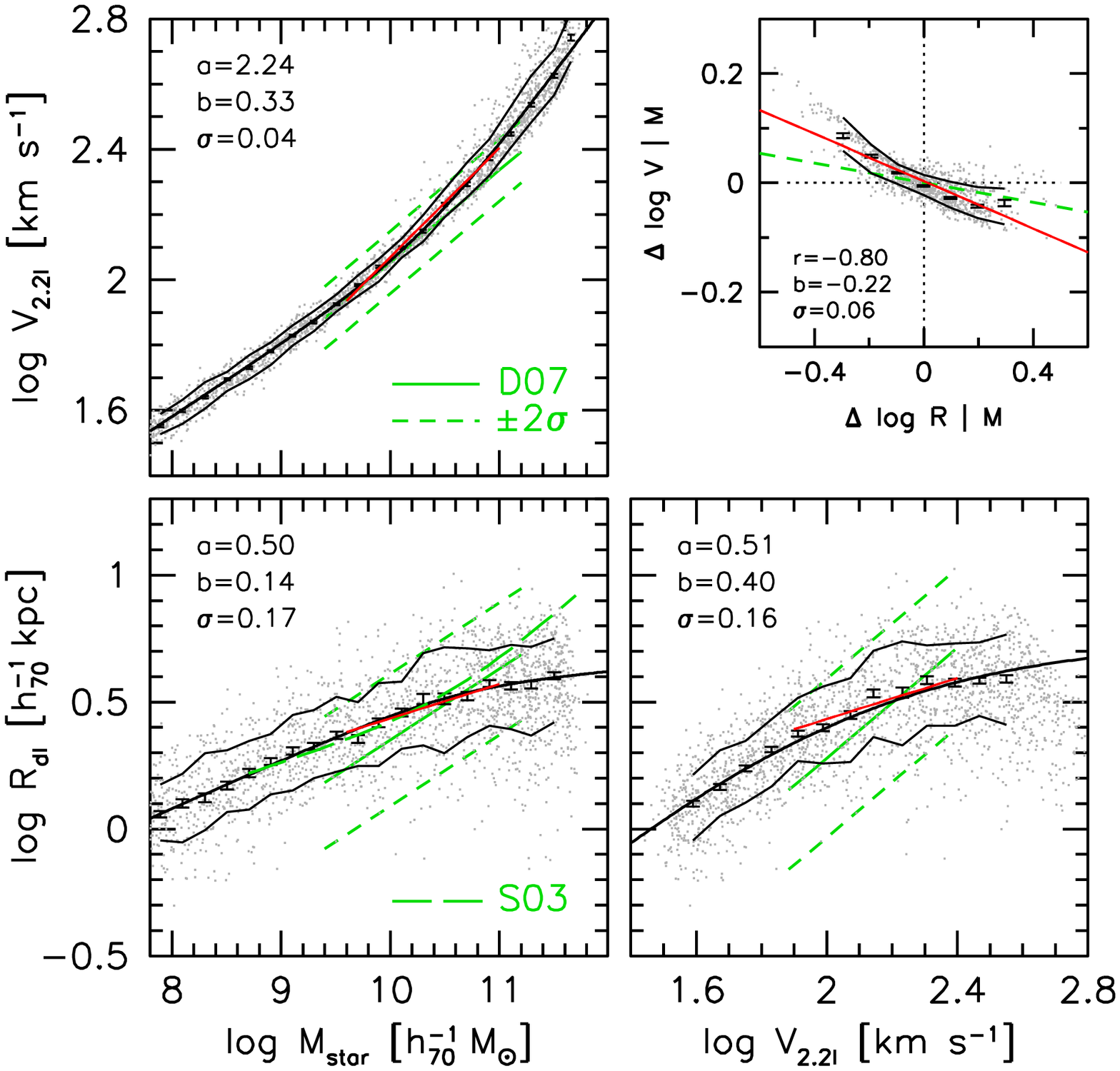,width=0.7\textwidth}}
\caption[]{\footnotesize As Fig.~\ref{fig:VMRId-nfb-nac} but for Model
  III (energy driven feedback). This model provides a reasonable match
  to the zero point, slope and scatter of the $VM$ relation. It
  predicts a shallower slope to the $RM$ and $RV$ relations than our
  fiducial data from Dutton \etal 2007, however at low masses the
  shallow slope of the model is in good agreement with the data from
  Shen \etal (2003) (green long dashed). The correlation between the
  residuals of the $VM$ and $RM$ relations (top right panel) of the
  models is stronger than the observations, although observational
  uncertainties in stellar mass measurements will cause the observed
  correlation to be underestimated.}
\label{fig:VMRId-efb-nac}
\end{figure*}

\subsection{Models with Scatter}
Having discussed the effects of  energy vs. momentum driven winds, and
halo contraction vs  no halo contraction for a  single halo mass, with
the median concentration, and angular momentum parameters, we now turn
our attention  to models with the  full range of  halo masses relevant
for  disk  galaxies,  and  with distributions  of  concentration,  and
angular momentum parameters.

We  run a  Monte  Carlo simulations  with  halo masses  ranging from  $
10^{10}  \lta \Mvir  \lta 10^{13}$  $\hMsun$, corresponding  to virial
velocities  ranging   from  $31  \lta   \Vvir  \lta  310   \kms$.   In
$\Lambda$CDM  there are  many  more  low mass  haloes  than high  mass
haloes,  however, since  we are  interested in  the  scaling relations
between galaxies, rather than the  number densities we sample the halo
masses uniformly in log-space.

As discussed in Dutton \etal (2007) we also find that models with the
expected  scatter in  halo  spin parameter  $\sigma_{\ln\lambda}=0.54$
significantly over predict the amount  of scatter in the $RM$ and $RV$
relations.   This may  signify that  disk galaxies  form in  a special
sub-set  of   haloes,  or  that   the  baryons  acquire   a  different
distribution of  specific angular momentum  than the dark  matter. For
the remainder  of this paper we adopt  $\sigma_{\ln \lambda}=0.35$, as
this provides a  reasonable agreement to the observed  scatter in disk
sizes.

To illustrate  the effect of  feedback on galaxy scaling  relations we
consider three models.  Model I has no feedback, Model II has momentum
feedback,  and Model  III  has  energy feedback  and  an average  spin
parameter a factor of 1.4 lower  than models I and II.  The parameters
of these  models are given in Table  \ref{tab:params}.  The parameters
of Model II and III were chosen  to match the zero points of the $VMR$
relations,  and  thus  for  reasons  discussed  above,  they  have  no
adiabatic contraction.  For each model we run a Monte-Carlo simulation
consisting of 2000 galaxies.  For each galaxy we select the parameters
$c$, $\lambda$, and $\alpha$  from log-normal distributions with means
and scatters as determined by the parameters in \S~\ref{sec:param} and
Table \ref{tab:params}.

Figs.~\ref{fig:VMRId-nfb-nac}-\ref{fig:VMRId-efb-nac}  show  the $VMR$
relations, as  well as  the correlation between  the residuals  of the
$VM$  and $RM$ relations,  for models  I-III. Recall  that $V$  is the
circular velocity measured at  $2.15$ $I$-band disk scale lengths, $M$
is the stellar mass, and $R$ is the $I$-band disk scale length.

\subsubsection{slopes}
The  VMR relations  in these  figures are  fit with  two  relations: a
single power-law over the range where there is observational data, and
a double power-law  over the full range of  masses.  The parameters of
the  best-fit single  power-law fits  are indicated  in the  panels of
Figs.~\ref{fig:VMRId-nfb-nac}-\ref{fig:VMRId-efb-nac}.    The   double
power-law is given by
\begin{equation}
  y = y_0\left(\frac{x}{x_0}\right)^{\alpha}\left[\frac{1}{2}+\frac{1}{2}\left(\frac{x}{x_0}\right)\right]^{(\beta-\alpha)}
\label{eq:power1}
\end{equation}
Here  $\alpha$ is  the  slope at  $x\ll x_0$;  $\beta$ is  the slope  at
$x \gg x_0$; $x_0$ is the transition scale;  $y_0$ is the value of $y$ at
$x_0$. The best
fit values of these parameters are given in Table~\ref{tab:power1}.

\begin{table}
 \centering
\begin{minipage}{0.4\textwidth}
  \caption{Parameters  of double  power-law fits  to VMR  relations in
    Figs.~\ref{fig:VMRId-nfb-nac}-\ref{fig:VMRId-efb-nac}   using  the
    function in Eq.~(\ref{eq:power1}) . }
  \begin{tabular}{cccccc}
\hline  
  Model &  $\alpha$ & $\beta$ & $M_{\rm star,0}$ & $V_{2.2I,0}$ \\
    I  & 0.27 &  0.34 & 10.71 & 2.34 \\
   II  & 0.29 &  0.44 & 10.94 & 2.36 \\
  III  & 0.22 &  0.46 & 10.35 & 2.17 \\
\\
  Model &  $\alpha$ & $\beta$ & $M_{\rm star,0}$ & $R_{\rm dI,0}$ \\
    I  &  0.44 & 0.14 & 11.56 & 0.65 \\
   II  &  0.28 & 0.17 & 11.15 & 0.68 \\
  III  &  0.19 & 0.04 & 10.55 & 0.52 \\
\\
  Model &  $\alpha$ & $\beta$ & $V_{\rm 2.2I,0}$ & $R_{\rm dI,0}$ \\
    I  &  1.79 &  0.40 & 2.59 & 0.65 \\
   II  &  1.15 &  0.00 & 2.50 & 0.72 \\
  III  &  1.15 &  0.00 & 2.00 & 0.40 \\
\hline 
\label{tab:power1}
\end{tabular}
\end{minipage}
\end{table}

The slope (as given by the single power-law fits) of the $VM$ relation
is  only weakly  dependent on  the feedback  model.  This  is expected
since (as  shown in Figs.~\ref{fig:VMRIm-efb}  \& \ref{fig:VMRIm-mfb})
the offset of a galaxy from the $VM$ relation is only weakly dependent
on  the galaxy  mass fraction,  which is  determined by  feedback (for
haloes with $\Mvir \lta 10^{12}\Msun$).   By contrast the slope of the
$RM$ relation depends strongly on  the feedback model.  Again, this is
expected  given that the  offset of  a galaxy  from the  $RM$ relation
depends  strongly  on the  galaxy  mass  fraction.  The model  without
feedback  (model I) has  a slope  of $0.40$,  the model  with momentum
driven feedback (model  II) has a slope of $0.26$,  and the model with
energy driven feedback (model III) has a slope of $0.14$.

The observed slope of the size-mass relation from (Courteau \etal 2007
and Dutton \etal 2007) is $\simeq 0.28$, which favors the momentum
driven wind model.  However, as discussed in \S~\ref{sec:obsvlr}, at
low stellar masses ($\Mstar \lta 10^{10}\Msun$), this slope is likely
biased high by selection effects.  Shen \etal (2003) find a slope of
0.14 at low masses for the half-light radius-stellar mass relation for
a much more complete sample of late-type galaxies. Such a shallow
slope is in much better agreement with our energy driven model.  Thus
the observed slope of the $RM$ relation favors the energy driven wind
model at low masses, and the momentum driven model at high
masses. However, at high stellar masses ($\Mstar \gta 10^{10.5}\Msun$)
bulges are common is spiral galaxies (e.g. Weinzirl \etal 2008).  Due
to the correlation between the masses of bulges and black holes
(Magorrian \etal 1998), AGN feedback may play a significant role in
regulating galaxy formation efficiency in high mass spiral galaxies.
Thus it is {\it plausible} that a model with energy driven SN feedback
(which primarily effects galaxies in low mass haloes) and AGN feedback
(which primarily effects galaxies in high mass haloes) could explain
the slopes of the size-mass relation at low and high masses. However,
since AGN feedback is not expected to be significant for galaxies in
low mass haloes, it is {\it unlikely} that AGN feedback will be able
to help the momentum driven wind model reproduce the shallow slope of
the size-mass relation at low masses.

\begin{figure*}
\centerline{
\psfig{figure=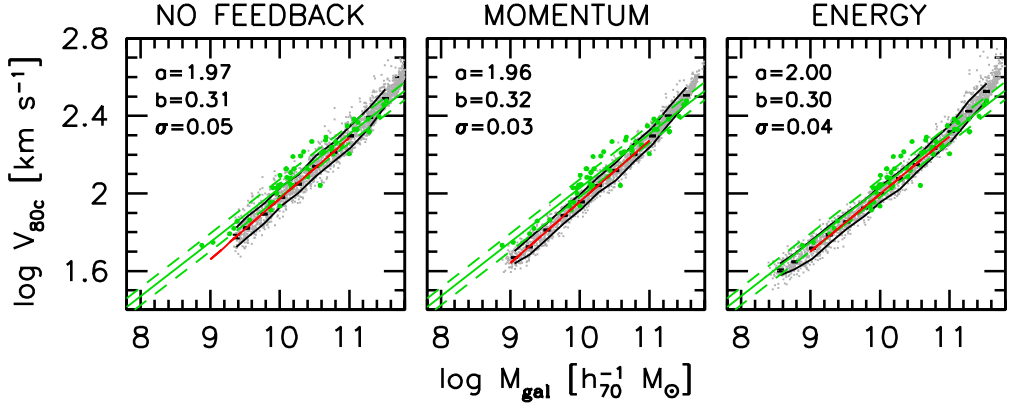,width=0.8\textwidth}}
\caption[]{\footnotesize Dependence of the Baryonic Tully-Fisher (BTF)
  relation at redshift $z=0$ on feedback.  The green points show data
  from McGaugh (2005), where the rotation velocities have been
  measured in the ``flat'' part of resolved HI rotation curves. Green
  lines show the mean and $1\sigma$ scatter of a fit to this data. The
  grey points show the models, where $V_{80c}$ is the circular
  velocity at a radius enclosing 80\% of the gas. The black lines show
  the 16th and 84th percentiles in bins of width 0.25 dex in
  $\Mgal$. The red lines show a fit of the form: $\log V_{80c} = a +
  b(\log\Mgal -10)$. The fits are performed over the range that the
  red lines are plotted. The zero point, slope, and scatter for each
  fit are given in the upper right corner of each panel. The slope,
  zero point and scatter of the BTF relation is only weakly dependent
  on the feedback model.}
\label{fig:btf}
\end{figure*}

\subsubsection{scatter and residual correlations}
All models produce a $VM$ relation with relatively small scatter, with
smaller scatter in the models with feedback.  The amount of scatter in
the $VM$ relation is directly related to the strength of the
correlation between the $VM$ and $RM$ relations. The model without
feedback has a very strong correlation (correlation coefficient,
$r=-0.98$; slope, $b=-0.39$), which is caused by these galaxies being
baryon dominated at 2.2 disk scale lengths. The models with feedback
have weaker, but still significant, correlations ($r\simeq-0.8$,
$b\simeq -0.2$).  These correlations are stronger than that observed
by Courteau \etal (2007) for the $I$-band $VL$ and $RL$ relations
($r=-0.16$, $b=-0.07$), and by Gnedin \etal (2007) ($r=0.23\pm0.14$)
and Avila-Reese \etal (2008) ($r=-0.29$, $b=-0.09$) for the stellar
mass $VM$ and $RM$ relations.  However, as discussed in Dutton \etal
(2007) scatter in the stellar mass-to-light ratio of $\simeq 0.15$
dex, either from intrinsic variations or measurement uncertainties
will weaken the correlation between the observed relations compared to
the theoretical $VM$ and $RM$ relations.  Thus we do not consider this
failure of the model as a serious shortcoming.

\subsection{The Baryonic Tully-Fisher Relation}
\label{sec:btf}
The fundamental basis of the Tully-Fisher (linewidth-luminosity)
relation is believed to be the relation between the asymptotic
rotation velocity of a galaxy disk, $V_{\rm flat}$, and the baryonic
mass, $M_{\rm gal}$, (the sum of stellar and cold gas mass). This
relation is referred to as the Baryonic Tully-Fisher (BTF)
relation. It was first studied by McGaugh \etal (2000), and
subsequently by Bell \& de Jong (2001), McGaugh (2005), Geha \etal
(2006), Noordermeer \& Verheijen (2007), and Avila-Reese \etal (2008).

The most significant source of uncertainty in the BTF is how one
measures stellar mass. McGaugh (2005) measured the BTF for stellar
masses calculated using different methods: stellar population
synthesis models (e.g. Bell \& de Jong 2001), the maximum disk
hypothesis (van Albada \& Sancisi 1986), the mass-discrepancy
acceleration relation (i.e. Modified Newtonian Dynamics, or MOND,
Milgrom 1983)).  McGaugh (2005) found that the scatter in the BTF was
minimized when the stellar masses were calculated with the
mass-discrepancy acceleration relation.  Under the assumption that the
correct method would minimized the scatter in the BTF, this is
evidence in favor of MOND over dark matter.

However, this is a circular argument because a relation between the
asymptotic rotation velocity of a galactic disk, $V_{\rm flat}$ and
the baryonic mass, $M_{\rm gal}$, with {\it zero scatter}, is built
into MOND.  Thus if the stellar masses are chosen based on the MOND
prescription, they will result in a BTF with scatter only due to
measurement errors on $V_{\rm flat}$ and distance uncertainties.  The
scatter in the BTF (as defined as the relation between $V_{\rm flat}$
and $M_{\rm gal}$) thus cannot be used to discriminate between MOND
and dark matter. However, MOND generally predicts higher stellar
masses than stellar population synthesis models (based on a Kroupa
IMF). Thus if stellar masses could be measured independently, this
would provide a means of falsifying MOND.

Here we use the data from McGaugh (2005), using the stellar population
synthesis stellar mass-to-light ratios, with an offset of -0.1 dex
(corresponding to a Chabrier IMF).  The majority of galaxies in this
sample are in the UMa Cluster, for which the distance is somewhat
uncertain. McGaugh (2005) adopted a distance of 15 Mpc. We adopt the
{\it HST} Key Project distance of $D=20.7$ Mpc (Sakai \etal 2000),
which is also the distance used by Bell \& de Jong (2001).

The BTF data are plotted as green filled circles in
Fig.~\ref{fig:btf}.  A linear fit to the data gives the following
relation between the rotation velocity and baryonic mass
\begin{equation}
\log \frac{V_{\rm flat}}{[\kms]} = 2.027+0.279\left(\log\frac{M_{\rm gal}}{[M_{\odot}]} -10\right)
\end{equation}
with a scatter of 0.053 dex in $\log V_{\rm flat}$. This BTF is
consistent with that from Bell \& de Jong (2001) who report a slope of
$0.285(\pm 0.015)$ and a zero point of $2.031(\pm 0.011)$, and that
from Avila-Reese \etal (2008) who report a slope of $0.306(\pm0.012)$
a zero point of $2.036(\pm 0.129)$ and an intrinsic scatter of 0.051
dex in $\log V$. The slightly steeper slope obtained by Avila-Reese
\etal (2008) can be attributed (see Verheijen 2001) to their use of
$H_{\rm I}$ linewidths, compared to $V_{\rm flat}$ used by McGaugh
(2005) and Bell \& de Jong (2001). This good agreement is reassuring
given that the data samples are largely based on the data set of
Verheijen \etal (2001).

The BTF relations for our models are given by the grey dots in
Fig.~\ref{fig:btf}. For the rotation velocity we use $V_{80c}$, the
circular velocity at a radius enclosing $80\%$ of the gas mass,
$R_{80c}$, which usually corresponds to the flat part of the rotation
curve (see Dutton 2008 rotation curves).  Power-law fits to the models
over baryonic masses between $1\times 10^9$ and $1\times 10^{11}$ are
shown as red lines in the figure. The parameters of these fits are
given in the top left of each panel.  All three models result in BTF
relations with similar slopes, zero points and scatter, and in
reasonable agreement with observations.  We note that for the model
galaxies the slope of the BTF depends on galaxy mass, with slightly
larger slopes for higher mass galaxies.

The BTF relation has been used to constrain the relation between
baryonic mass and halo mass.  By comparing the observed slope of the
BTF ($0.27\pm0.01$) to the prediction from CDM (the slope of the
$\Vmaxh-\Mvir$ relation for dark matter haloes is $0.294\pm0.005$,
Bullock \etal 2001a), Geha \etal (2006) argued that low mass galaxies
have not preferentially lost baryons as would be predicted by feedback
models (e.g. Dekel \& Silk 1986). However, this is based on the
incorrect assumption that the maximum observed rotation velocity is
equal (or proportional) to the maximum circular velocity of the halo,
{\it independent} of the baryon to halo mass ratio.  As discussed by
several authors (e.g. Navarro \& Steinmetz 2000; Dutton \etal 2007;
Avila-Reese \etal 2008), the maximum rotation velocity of a galaxy is,
in general, not equal to the maximum circular velocity of the halo in
the absence of galaxy formation, $\Vmaxh$. As the baryon fraction
increases, so to does the maximum circular velocity. This is because
the baryons contribute a non-negligible amount of mass to the circular
velocity. Thus for reasonable galaxy mass fractions, variation in
galaxy mass fraction moves galaxies roughly parallel to the BTF.

Fig.~\ref{fig:btf} shows that the slope, zero point and scatter of the
BTF are remarkably insensitive to the feedback model.  Furthermore, as
shown in \S 4.1, our energy feedback model results in substantial
differential mass loss between haloes of mass $\Mvir \simeq 10^{10}
\Msun$ and $\Mvir \simeq 10^{12}\Msun$. Yet it has the same BTF slope
as a model with no mass loss (and constant baryon to dark matter ratio
within this range of halo masses).  This provides a counter example to
the claim by Geha \etal (2006) that models with preferential mass loss
in dwarf galaxies cannot explain the slope of the BTF.

\begin{figure*}
\centerline{
\psfig{figure=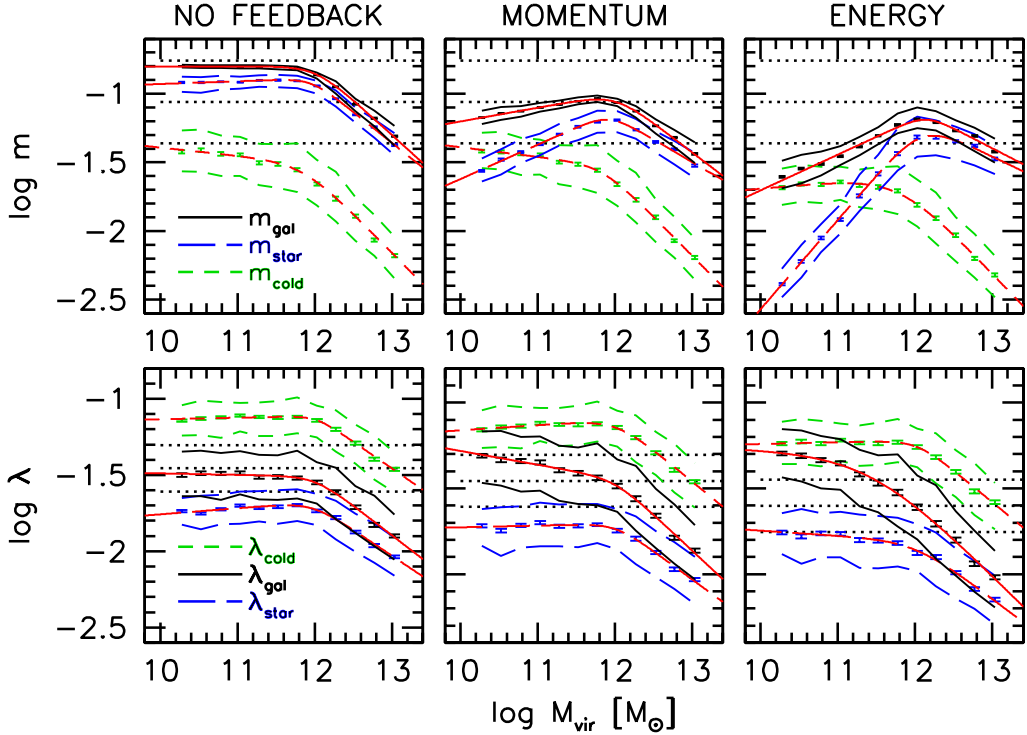,width=0.8\textwidth}}
\caption[]{\footnotesize Dependence of mass fractions and spin
  parameters on virial mass for our 3 feedback models: no feedback
  (left); momentum feedback (middle); energy feedback (right).  The
  lines show the 16th and 84th percentiles in bins of width 0.25 dex
  in $\Mvir$, the error bars show the Poisson error on the median.
  The upper panels show the mass fractions of the galaxy,
  $\mgal=\Mgal/\Mvir$ (black, solid), stars, $\mstar=\Mstar/\Mvir$
  (blue, long-dashed), and cold gas $\mcold=M_{\rm cold}/\Mvir$
  (green, short-dashed).  The dotted lines show the cosmic baryon
  fraction, $f_{\rm bar}$, as well as $f_{\rm bar}/2$, and $f_{\rm
    bar}/4$.  The red lines show double power-law fits to the medians,
  the parameters of these fits are given in Table~\ref{tab:mgalfit}.
  The lower panels show the spin parameters of the galaxy, $\lamgal =
  (\jgal/\mgal)\lambda$, (black, solid), stars,
  $\lamstar=(\jstar/\mstar)\lambda$, (blue, long-dashed), cold gas
  $\lamcold=(j_{\rm cold}/m_{\rm cold})\lambda$ (green, short-dashed),
  and dark matter, $\lambda $, (black, dotted).}
\label{fig:mglg3}
\end{figure*}

\section{The dependence  of galaxy mass fractions  and spin 
parameters  on feedback}
\label{sec:mgal}
The two main parameters that  determine the structure of disk galaxies
are the mass and specific angular  momentum of the cold gas and stars.
We now investigate the dependence  of these parameters on feedback and
halo mass. 

\begin{table}
 \centering
\begin{minipage}{0.40\textwidth}
  \caption{Parameters of  double power-law fits to  mass fractions and
    spin parameters  vs virial mass in  Fig.~\ref{fig:mglg3} using the
    function in Eq.~(\ref{eq:power2}) . }
  \begin{tabular}{cccccc}
\hline  
  Model &  $\alpha$ & $\beta$ & $\gamma$ &  $M_{\rm vir,0}$ & $m_{\rm cold,0}$ \\
    I  & -0.06 &  -0.57 &  2.00 & 11.85 & -1.58 \\
   II  & -0.07 &  -0.57 &  2.45 & 11.84 & -1.58 \\
  III  &  0.05 &  -0.59 &  1.50 & 11.80 & -1.73 \\
\\
  Model &  $\alpha$ & $\beta$ & $\gamma$ &  $M_{\rm vir,0}$ & $m_{\rm star,0}$ \\
    I  &  0.02 &  -0.45 &  3.54 & 11.96 & -0.93 \\
   II  &  0.25 &  -0.33 &  3.60 & 11.92 & -1.19 \\
  III  &  0.65 &  -0.27 &  2.55 & 12.09 & -1.32 \\
\\
  Model &  $\alpha$ & $\beta$ & $\gamma$ &  $M_{\rm vir,0}$ & $m_{\rm gal,0}$ \\
    I  &  0.00 &  -0.51 &  2.59 & 12.02 & -0.86 \\
   II  &  0.10 &  -0.43 &  2.71 & 12.01 & -1.06 \\
  III  &  0.26 &  -0.29 &  3.90 & 12.14 & -1.19 \\
\\
  Model &  $\alpha$ & $\beta$ & $\gamma$ &  $M_{\rm vir,0}$ & $\lambda_{\rm cold,0}$ \\
    I  &  0.01 &  -0.35 &  4.14 & 12.02 & -1.14 \\
   II  &  0.03 &  -0.37 &  2.45 & 12.03 & -1.15 \\
  III  &  0.01 &  -0.37 &  2.95 & 12.03 & -1.26 \\
\\
  Model &  $\alpha$ & $\beta$ & $\gamma$ &  $M_{\rm vir,0}$ & $\lambda_{\rm star,0}$ \\
    I  &  0.04 &  -0.34 &  2.79 & 11.97 & -1.72 \\
   II  &  0.01 &  -0.33 &  3.10 & 12.02 & -1.74 \\
  III  & -0.03 &  -0.34 &  1.75 & 12.01 & -1.86 \\
\\
  Model &  $\alpha$ & $\beta$ & $\gamma$ &  $M_{\rm vir,0}$ & $\lambda_{\rm gal,0}$ \\
    I  & -0.01 &  -0.39 &  2.80 & 12.01 & -1.55 \\
   II  & -0.08 &  -0.42 &  2.70 & 12.00 & -1.48 \\
  III  & -0.02 &  -0.45 &  0.95 & 11.44 & -1.44 \\
\hline 
\label{tab:mgalfit}
\end{tabular}
\end{minipage}
\end{table}

\subsection{Galaxy Mass Fractions}
Fig.~\ref{fig:mglg3} shows  the mass  fractions and spin  parameters of
our 3  models as a function  of virial mass. 

The relations  in Fig.~\ref{fig:mglg3}  are fitted with  the following
double power-law:
\begin{equation}
  y = y_0\left(\frac{x}{x_0}\right)^{\alpha}\left[\frac{1}{2}+\frac{1}{2}\left(\frac{x}{x_0}\right)^\gamma\right]^{(\beta-\alpha)/\gamma}
\label{eq:power2}
\end{equation}
Here  $\alpha$ is  the  slope at  $x \ll x_0$;  $\beta$ is  the slope  at
$x \gg x_0$; $x_0$ is the transition scale;  $y_0$ is the value of $y$ at
$x_0$; and  $\gamma$ determines how  fast the transition is.  The best
fit values of these parameters are given in Table~\ref{tab:mgalfit}.

We  start our discussion  with model  I, which  has no  feedback (left
panels).   For low  mass haloes  the galaxy  mass fraction  $\mgal$ is
close to  the universal baryon fraction, $\fbar\simeq  0.17$.  This is
because cooling  is very efficient in  low mass haloes.   Above a halo
mass of  $\Mvir \simeq 10^{12}  \Msun$ the galaxy mass  fraction drops
significantly, due to the inefficiency of cooling in high mass haloes.
We hereafter refer to this mass scale as the cooling threshold.

The  effect of feedback  is to  remove cold  gas from  the galaxy-halo
system. The  efficiency with which  feedback can eject gas  depends on
both the  depth of the potential  well, the amount  of star formation,
and on the wind model. The net effect in both energy and momentum wind
models  is for mass  to be  lost preferentially  in lower  mass haloes
(i.e.   potential well  dominates  over star  formation efficiency  at
fixed halo  concentration and angular momentum  parameters).  Note the
fraction of mass lost varies  smoothly with virial mass, i.e. there is
no  threshold for  mass  loss, as  for  example there  would  be in  a
constant wind velocity model. However, the energy driven wind model is
much  more efficient  at  removing baryons  from  haloes below  $\Mvir
\simeq 10^{12}\Msun$. This results  in very different scalings between
$\mgal$ and  $\mstar$ with  $\Mvir$ for the  two feedback  models. The
parameters of these scalings are given in Table~\ref{tab:mgalfit}.  In
principle  these  differences  are  testable with  galaxy-galaxy  weak
lensing  and/or satellite  kinematics measurements  of  virial masses,
combined with measurements of  stellar masses and neutral hydrogen gas
masses.   For  both  feedback  models  the  maximum  galaxy  formation
efficiency (defined  as $\mgal/\fbar$) occurs around a  virial mass of
$\Mvir  \simeq 10^{12}\Msun$, and  is $\simeq  0.50$ for  the momentum
driven wind and $\simeq 0.35$ for the energy driven winds.

Fig.\ref{fig:mglg3} also  shows the  cold gas mass  fraction $\mcold$,
which is just  the difference between $\mgal$ and  $\mstar$. Below the
cooling threshold,  $\mcold$ is  almost independent of  $\Mvir$, while
above the cooling threshold  $\mcold$ strongly decreases with $\Mvir$.
These trends  of $\mcold$ with  $\Mvir$ are qualitatively  similar for
all three feedback models. Thus feedback has a much stronger impact on
the  stellar  mass fraction,  $\mstar$,  than  on  the cold  gas  mass
fraction, $\mcold$.

\begin{figure*}
\centerline{
\psfig{figure=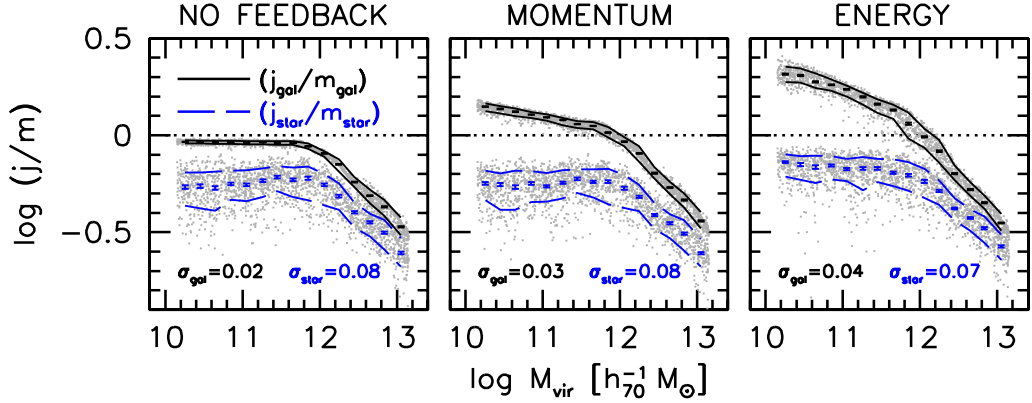,width=0.8\textwidth}}
\caption[]{\footnotesize Dependence of ratio between the specific
  angular momentum of the baryons/stars (black/blue) and dark matter
  on virial mass for our 3 models.  The lines show the 16th and 84th
  percentiles in bins of width 0.25 dex in $\Mvir$. The scatter about
  the median relation are given by $\sigma_{\rm gal}$ for
  $(\jgal/\mgal)$, and $\sigma_{\rm star}$ for $(\jstar/\mstar)$.  The
  dotted line corresponds to the specific angular momentum of the
  baryons and stars being equal to that of the dark matter.}
\label{fig:jm3}
\end{figure*}

\subsection{Galaxy Spin Parameters}

The assumption  that $(\jgal/\mgal)=1$ underlies  most applications of
the  Mo,  Mao, \&  White  (1998) disk  structure  models,  as well  as
observational attempts  at measuring the halo spin  parameter (e.g. van
den Bosch,  Burkert \&  Swaters 2001). Thus  an important  question is
whether  this assumption  is  valid in  galaxy  formation models  that
include inflows and outflows.

The lower  panels of Fig.~\ref{fig:mglg3} show the  spin parameters of
our 3 models as a function  of virial mass. We first discuss the model
without  feedback.  For haloes  below the  cooling threshold  the spin
parameter of the galaxy, $\lamgal$, is  almost the same as that of the
dark matter, which  is because almost all of  the baryons have cooled,
and thus  they bring in almost  all of the  angular momentum. However,
for haloes above the cooling threshold, $\lamgal$ decreases by about a
factor of  2 from $\Mvir=10^{12}$ to $10^{13}\Msun$.   This is because
the highest angular momentum material virializes at low redshifts, and
this gas does not have time to cool.  

As would be expected the stellar mass fraction and spin parameters are
lower  than the corresponding  parameters for  the total  galaxy.  The
lower stellar spin parameters are due to the star formation law, which
causes stars to  form less efficiently at larger  radii (where the gas
density  is lowest).   Thus the  gas disk  is more  extended  that the
stellar disk.

An interesting result  of our models with feedback  is that the galaxy
spin parameter can be higher than  the halo spin parameter. This is 
because  mass can  only  be ejected  from  radii where  there is  star
formation,  and  these  are  biased  towards small  radii,  where  the
specific angular  momentum of the gas  is lower than  the average.  As
shown  below  (in section  $\ref{sec:dmg}$),  at  a  given halo  mass,
feedback  is more efficient  at removing  baryons from  higher surface
density disks. Thus the increased star formation efficiency over comes
the deeper  potential well.  Note that  the scatter in  the galaxy and
stellar spin  parameters is roughly equal  to the scatter  in the halo
spin parameter, and this does not change with virial mass.

Although the  spin parameters  of the galaxy  and stars  are typically
different  from that of  the halo,  Fig.~\ref{fig:jm3} shows  that the
ratio of the  specific angular momentum of the baryons  to that of the
dark matter, $(\jgal/\mgal)$,  at a given virial mass  has very little
scatter (0.01 dex for model I, 0.02  dex for model II, and 0.04 dex of
model III).   The small  scatter holds for  all halo masses,  even for
galaxies that  have lost more than  half of their  baryons.  The small
scatter is  related to the fact  that there is very  little scatter in
the baryon mass fraction at a given halo mass.  By contrast there is a
much larger scatter (0.07-0.08 dex)  in the ratio between the specific
angular momentum of the stars and the dark matter.  This is due to the
dependence  of  global star  formation  efficiency  on galaxy  surface
density.

Given that the assumption that $(\jgal/\mgal)=1$ is violated in all of
our  models,  care  should  be  taken  when  interpreting  models  and
observational results based on this assumption. However, in our models
there is very little scatter  in $(\jgal/\mgal)$ at a fixed halo mass.
This suggests that the scatter  in the galaxy spin parameter (which is
in principle  observable), at  a given halo  mass, may be  an accurate
reflection  of the  scatter in  the halo  spin parameter  (which  is a
prediction of $\LCDM$, but not directly observable).

\begin{figure*}
\centerline{
\psfig{figure=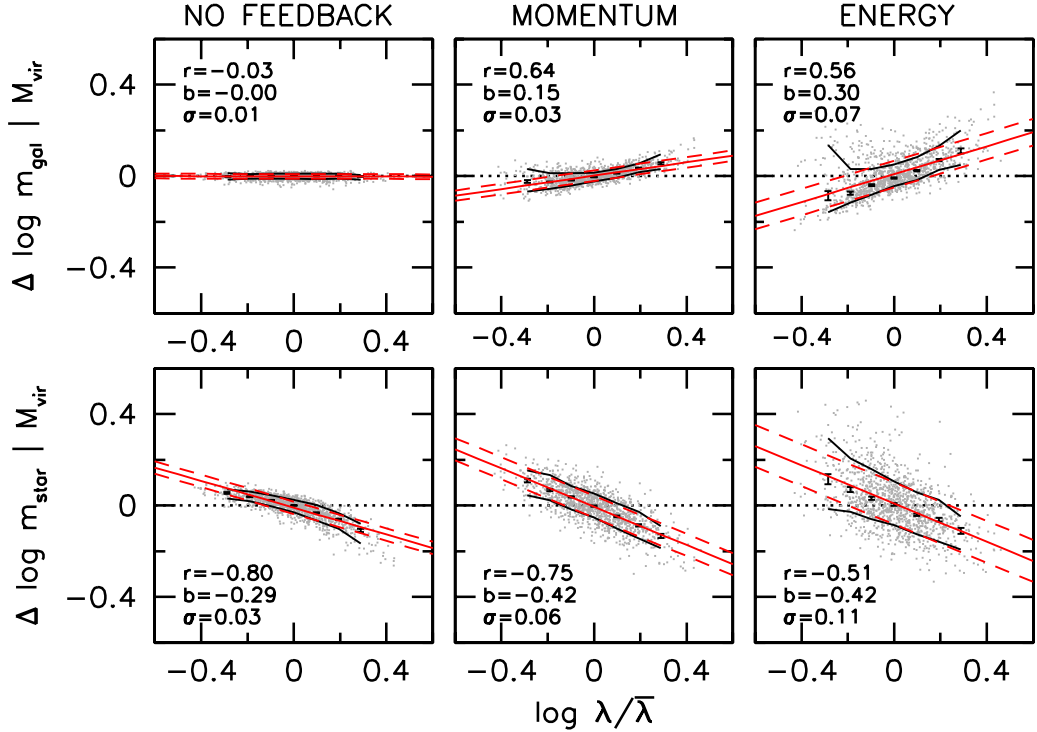,width=0.8\textwidth}}
\caption[]{\footnotesize Correlation between  scatter (at a given halo
  mass) in  halo spin, with scatter  (at a given halo  mass) in galaxy
  mass  fraction  (upper panels),  and  stellar  mass fraction  (lower
  panels).   The   black  lines  show  the  median,   16th,  and  84th
  percentiles of the models.  The  scatter in the galaxy mass fraction
  is correlated  with halo spin  (i.e.  feedback is more  efficient in
  higher  density  galaxies),  whereas  the scatter  in  stellar  mass
  fractions is anti-correlated with  halo spin (i.e. star formation is
  more efficient in higher density galaxies).  The red lines shows the
  mean (solid)  and 1$\sigma$ scatter (dashed)  of a fit  of the form:
  $\Delta\log m  = b\,  \log \lambda/\bar{\lambda}$.  The  slope, $b$,
  scatter,  $\sigma$  (in  $\log_{10}$  units),  and  the  correlation
  coefficient,  $r$,  are  given   for  each  fit.   The  correlation
  coefficient is  given by $r=b \,\sigma_x /  \sigma_y$, where $x=\log
  \lambda/\bar{\lambda}$ and $y=\Delta \log m | \Mvir$}
\label{fig:dmg3}
\end{figure*}

\subsection{Is Galaxy Mass Fraction Correlated with Halo Spin? }
\label{sec:dmg}

Fig.~\ref{fig:mglg3} shows that at a given virial mass there is only a
small scatter in  galaxy or stellar mass fractions.   For the momentum
driven wind model the scatter  in $\mgal$ and $\mstar$ is $\simeq 8\%$
and $\simeq 20\%$  respectively. For the energy driven  wind model the
scatters  are $\simeq 20\%$  and $\simeq  33\%$ respectively.  

We now  turn to the  question of where  this scatter comes  from.  The
upper  panels  of  Fig.~\ref{fig:dmg3}   show  the  residuals  of  the
$\mgal-\Mvir$ relations  vs scatter in  the halo spin  parameter.  For
the  no feedback  model,  the  baryon fraction  is  determined by  the
efficiency of  cooling.  In  our model, the  cooling efficiency  is to
first order  determined by the halo  mass.  The very  small scatter in
$\mgal$ results from scatter  in the halo concentration, which effects
the cooling  in two  ways. The halo  concentration determines  the MAH
(i.e.  low concentration haloes collapse later, and thus there is less
time  for the  baryons to  cool). The  halo concentration  effects the
density of the hot gas,  which directly effects the cooling time.  
For  the  energy  and  momentum   wind  models  there  is  a  positive
correlation ($r\simeq  0.6$) between $\mgal$ and $\lambda$  at a given
$\Mvir$.  Thus at a given virial mass, galaxies that have lower spins,
(i.e.  higher  surface density disks), are more  efficient at removing
baryons, despite the deeper potential well.

The  lower panels  of Fig.~\ref{fig:dmg3}  show the  residuals  of the
$\mstar-\Mvir$ relations.  All models show an anti-correlation between
the residuals, i.e.  galaxies that  form in lower spin haloes are more
efficient at  turning their  cold baryons into  stars. This  effect is
expected from the density dependent star formation recipe we adopt.

These results  have implications for  the scatter in  the Tully-Fisher
relation.  As discussed in \S 1,  one of the surprising aspects of the
Tully-Fisher relation  is that the  scatter is independent  of surface
brightness,  or  equivalently the  scatter  in  the  $VL$ relation  is
independent  of  the  scatter   in  the  $RL$  relation.   Firmani  \&
Avila-Reese (2000), van den Bosch (2000), and Dutton \etal (2007) have
shown that this could partially  be explained by the dependence of gas
fractions on  surface density, such  that lower surface  density disks
have higher  gas fractions.  Gnedin  \etal (2007), on the  other hand,
invoked  a correlation  between disk  mass fraction  and  disk surface
density to reduce the correlation between the $VM$ and $RM$ relations.
This works  as follows.  At a given  stellar mass there is  a range in
disk  sizes.  Smaller disks  should result  in larger  $V_{2.2}$, both
because the baryons contribute more to $V_{2.2}$, and because the halo
contribution  increases  due to  halo  contraction.   However, if  the
smaller disks live in lower mass haloes, then the reduced contribution
of  the  halo compensates  for  the  increased  contribution from  the
disk. This should result in a negative correlation between $\mgal$ and
$\lambda$,  opposite to  what we  find  in our  models.  Gnedin  \etal
(2007)  speculated that  feedback would  be less  efficient  in higher
surface  density  disks,  presumably  because the  potential  well  is
deeper.  However, in our models, the  reverse is the case.  At a given
halo mass higher surface density  disks are more efficient at removing
baryons because there  is more energy (or momentum)  input from SN due
to the higher star formation rates.

\subsection{The Dependence of Gas Fractions on Feedback}
\label{sec:fgas}
\begin{figure*}
\centerline{
\psfig{figure=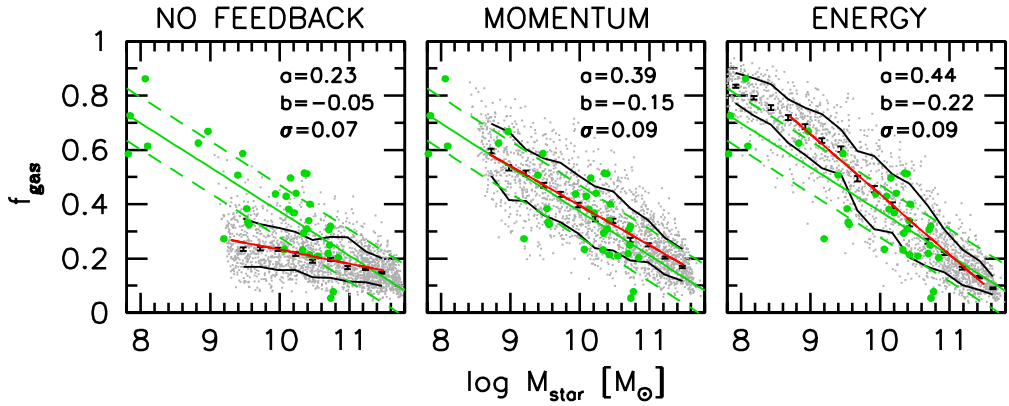,width=0.8\textwidth}}
\caption[]{\footnotesize Dependence of the gas fraction - stellar mass
  relation at redshift $z=0$ on feedback.  The green points show data
  from Garnett (2002), with green lines showing the mean and $1\sigma$
  scatter. The grey points show the models. The black lines show the
  16th and 84th percentiles in bins of width 0.25 dex in $\Mstar$. The
  red lines show a fit of the form: $f_{\rm gas} = a + b(\log\Mstar
  -10)$. The fits are performed over the range that the red lines are
  plotted. The zero point, slope, and scatter for each fit are given
  in the upper right corner of each panel. Note that models and data
  include both atomic and molecular gas. The model without feedback
  under predicts the gas fractions, whereas the models with feedback
  provide good matches to the observations.}
\label{fig:fgas}
\end{figure*}

Observationally it is known that the cold gas fraction is higher in
lower mass galaxies (e.g. McGaugh 1997 \& de Blok; Kannappan 2004;
Avila-Reese \etal 2008). Here we use data from Garnett 2002, who
compiled B-band magnitudes, B-V colors, atomic gas mass and molecular
gas mass for 31 spiral galaxies and 13 irregular galaxies. We compute
stellar masses using the following relation from Bell \etal (2003):
$\log (M_{\rm star}/L_B) = -0.941 +1.737(B-V) -0.1$, where the -0.1
corresponds to a Chabrier IMF. The gas fractions, defined as $f_{\rm
  gas} = M_{\rm cold}/(M_{\rm cold} + M_{\rm star})$, versus stellar
masses thus derived are plotted in green in Fig.~\ref{fig:fgas}.  A
linear fit gives the following relation between the gas fraction and
logarithm of stellar mass:
\begin{equation}
f_{\rm gas} = 0.374 -0.162 \left(\log \frac{M_{\rm star}}{[h_{70}^{-2} M_{\odot}]} -10 \right)
\end{equation}
with a scatter of 0.11 in $f_{\rm gas}$. Note that the ratio between
atomic and molecular is a strong function of stellar mass. Massive
galaxies have roughly equal amounts of atomic and molecular gas, while
the molecular gas fraction is negligible in galaxies with $M_{\rm
  star} \lta 10^{10}\Msun$. Thus ignoring the molecular gas
significantly underestimates the gas fractions at the high mass end.

Fig.~\ref{fig:fgas} shows the gas fraction vs stellar mass relation
for our three models at redshift $z=0$. The models are given by the
grey points. The model with no feedback (left panel) results in gas
fractions that are too low compared to the observations (green points
and lines). This is a result of the disks being too high density,
which causes star formation to proceed too fast.  Both models with
feedback result in gas fractions that are in good agreement with
observations (slope, zero point and scatter). In particular, the
momentum driven wind model is almost indistinguishable from the
observations. Distinguishing between these two feedback models would
require a more complete sample of galaxies with robust atomic and
molecular gas masses. 

\begin{figure*}
\centerline{
\psfig{figure=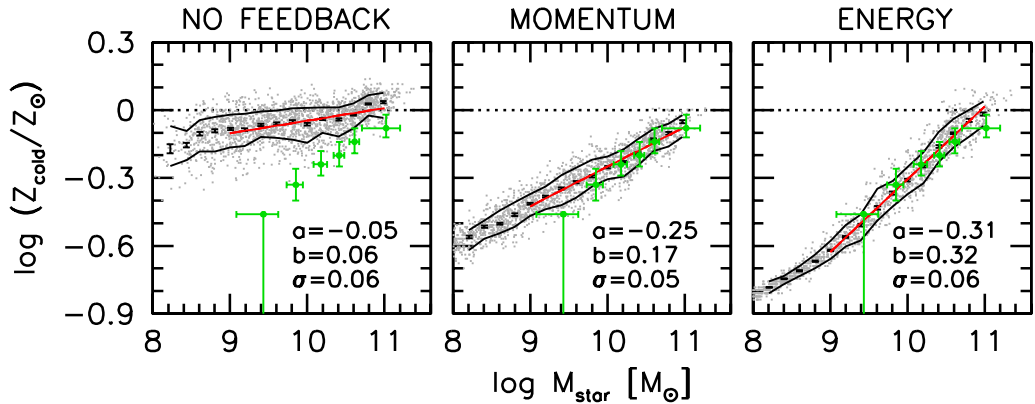,width=0.8\textwidth}}
\caption[]{\footnotesize Dependence of the Mass Metallicity Relation
  at $z=2.26$ on feedback. The black lines show the 16th and 84th
  percentiles of metallicity in bins of width 0.25 dex in
  $\Mstar$. The red lines show a fit of the form: $\log (Z_{\rm
    cold}/Z_{\odot}) = a + b(\log\Mstar -10)$.  The zero point, slope,
  and scatter for each fit are given in the lower right corner of each
  panel.  The green points show observational data at $z=2.26\pm0.17$
  from Erb \etal (2006). The model without feedback over predicts the
  metallicities, whereas the models with feedback provide good matches
  to the observations.}
\label{fig:zm3}
\end{figure*}

\begin{figure*}
\centerline{
\psfig{figure=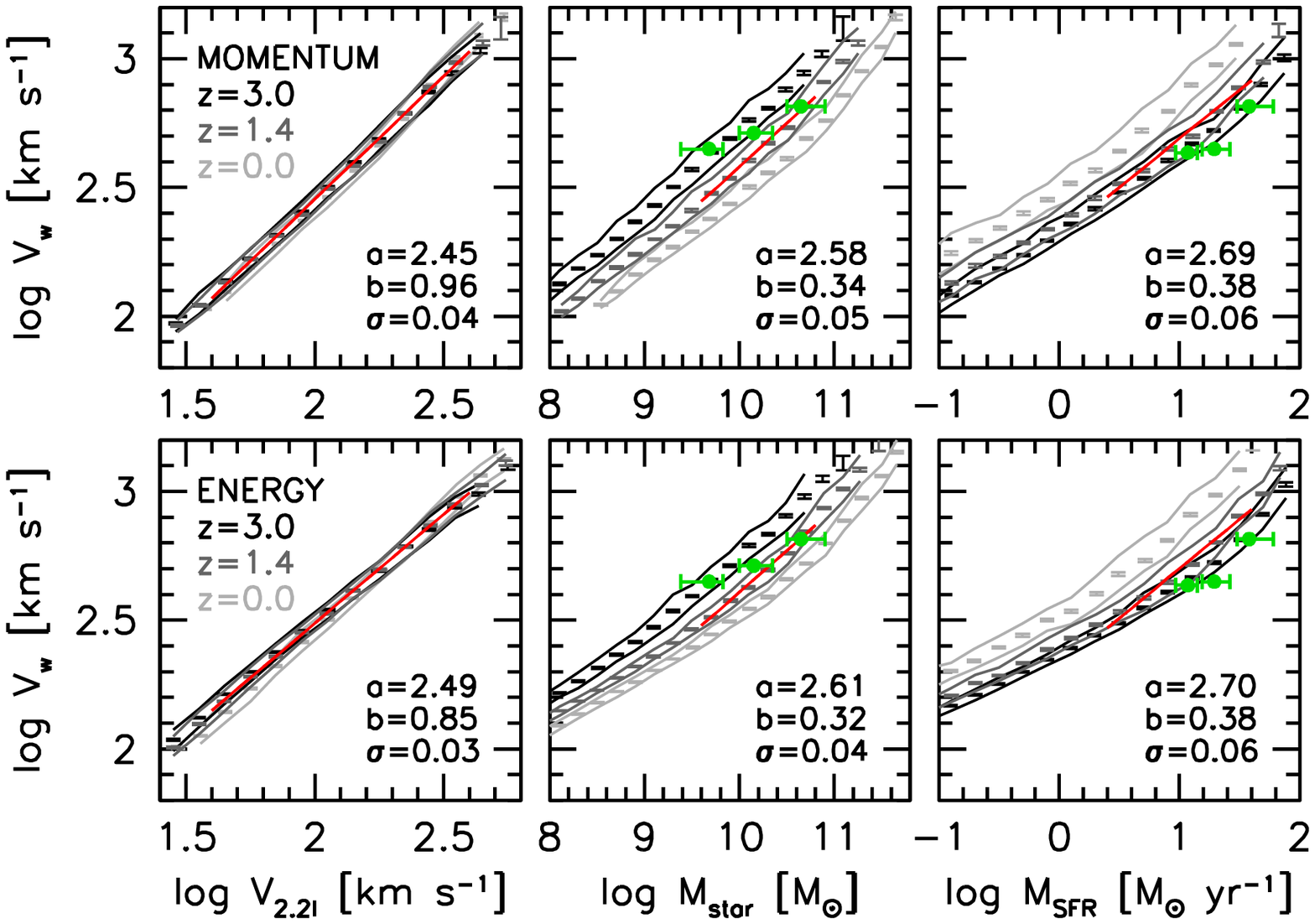,width=0.8\textwidth}}
\caption[]{\footnotesize Mass weighted wind velocity, $V_{\rm w}$,
  versus galaxy rotation velocity, $V_{2.2I}$, stellar mass, $\Mstar$,
  and star formation rate, $\dot {M}_{\rm star}$, for energy driven
  (lower panels) and momentum driven (upper panels) wind models.  The
  black and grey lines show the 16th and 84th percentiles of the models
  at $z=0.0$ (light grey), $z=1.4$ (dark grey) and $z=3.0$
  (black). The red lines show fits to the $z=1.4$ models. The fits are
  of the form: $\log V_{\rm w} = a + b (\log V_{2.2I}-2)$, $\log
  V_{\rm w} = a + b (\log M_{\rm star} -10)$, and $\log V_{\rm w} = a
  + b (\log M_{\rm SFR} -1)$. The slopes ($b$), zero points ($a$) and
  scatter $(\sigma)$ of these fits are given in the lower right
  corner of each panel.  The green points with error bars show
  observational data at $z\simeq 1.4$ from Weiner \etal (2009), see
  text for further details.  Both energy and momentum driven wind
  models predict similar scalings, that are broadly in agreement with
  the observations.}
\label{fig:vw3}
\end{figure*}


\section{The Dependence of the Mass Metallicity Relation on Feedback}
\label{sec:mz}
The relation between stellar mass (or luminosity) and gas (or stellar)
phase metallicity has long been thought to hold important clues to the
nature of galaxy  outflows. Indeed, the slope of  the mass metallicity
relation at  $z\simeq 2$ has been  used to argue in  favor of momentum
driven outflows over energy  driven outflows and no outflows (Finlator
\& Dav\'e  2008).  However,  the energy driven  wind model  adopted in
Finlator  \& Dav\'e  (2008) assumed  a constant  mass  loading factor,
$\eta=2$, and a constant wind velocity $V_{\rm w}=484 \kms$.  As noted
by Finlator  \& Dav\'e  (2008), this constant  wind model is  only one
possible   implementation  of   energy  driven   outflows.    Thus  an
interesting question  is whether the energy driven  outflow model that
we  have  implemented is  able  to reproduce  the  slope  of the  mass
metallicity relation at $z\simeq 2$.

Fig.~\ref{fig:zm3} shows the stellar mass-metallicity relation for our
three models at  redshift $z=2.26$.  The metallicity used  here is the
metallicity of the cold gas, which  is calculated as the sum of metals
in the  cold gas divided by  the mass of  cold gas. The model  with no
feedback has  a mass-metallicity relation  with very shallow  slope of
$b\simeq  0.06$, and a  mean metallicity  close to  that of  the yield
(i.e. Solar). The green points  with error bars show the observational
results for  the gas phase oxygen  abundance vs stellar  mass from Erb
\etal   (2006),   assuming   a   solar   oxygen   abundance   of   $12
+\log(O/H)=8.66$.   Our no  feedback  model is  inconsistent with  the
observations, both in terms of  slope and zero point.  The models with
feedback result  in steeper slopes of  the mass-metallicity relations:
$b\simeq 0.17$ for the momentum  wind model and $b\simeq 0.32$ for the
energy wind model.  Both of  these models are in qualitative agreement
with  the observations.  Overall  the energy  driven model  provides a
better  match  to  the   observed  slope,  but  given  the  systematic
uncertainties  in metallicity  measurements it  would be  premature to
strongly distinguish between the energy and momentum outflow models.

Finally  we note that  all of  our models  result in  mass metallicity
relations with small scatter $\simeq  0.06$ dex. Thus we conclude that
outflows   are  not  responsible   for  the   small  scatter   in  the
mass-metallicity relation.  The origin of this small  scatter, as well
as the evolution of the mass metallicity relation will be discussed in
a future paper.


\section{How do galaxies lose their gas? }
\label{sec:mdot}
Having  established that  in  models with  feedback,  galaxies lose  a
significant  fraction of  their accreted  baryons, we  now investigate
when   and  how   the  baryons   are   lost.   The   left  panels   of
Fig.~\ref{fig:vw3} show the mass  weighted wind velocity, $V_{\rm w}$,
versus the  rotation velocity at  2.2 disk scale  lengths, $V_{2.2I}$,
for models at redshifts, $z=0.0,  1.4,$ and $3.0$. This relation has a
slope $\simeq 1$, small scatter, and holds at all redshifts.  For both
energy and  momentum driven  winds the mean  wind velocity  is roughly
2.5-3 times $V_{2.2I}$.  However,  this relation is no surprise, since
by construction we assume that the local wind velocity is equal to the
local escape  velocity.  The non-zero scatter in  the relation between
$V_{\rm w}$ and $V_{2.2I}$ is due to the fact that, in our model, mass
is  ejected  from a  range  of  radii, and  hence  a  range of  escape
velocities.

\begin{figure}
\centerline{
\psfig{figure=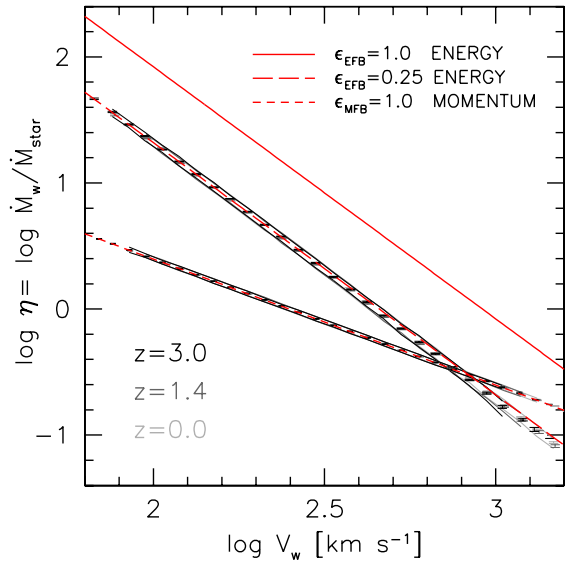,width=0.45\textwidth}}
\caption[]{\footnotesize Mass loading factor versus mean (outflow mass
  weighted)  wind velocity  for our  energy and  momentum  driven wind
  models at redshifts $z=0.0, 1.4$ and  $3.0$. The red lines are as in
  Fig.~\ref{fig:etav}.}
\label{fig:etav2}
\end{figure}

\begin{figure*}
\centerline{
\psfig{figure=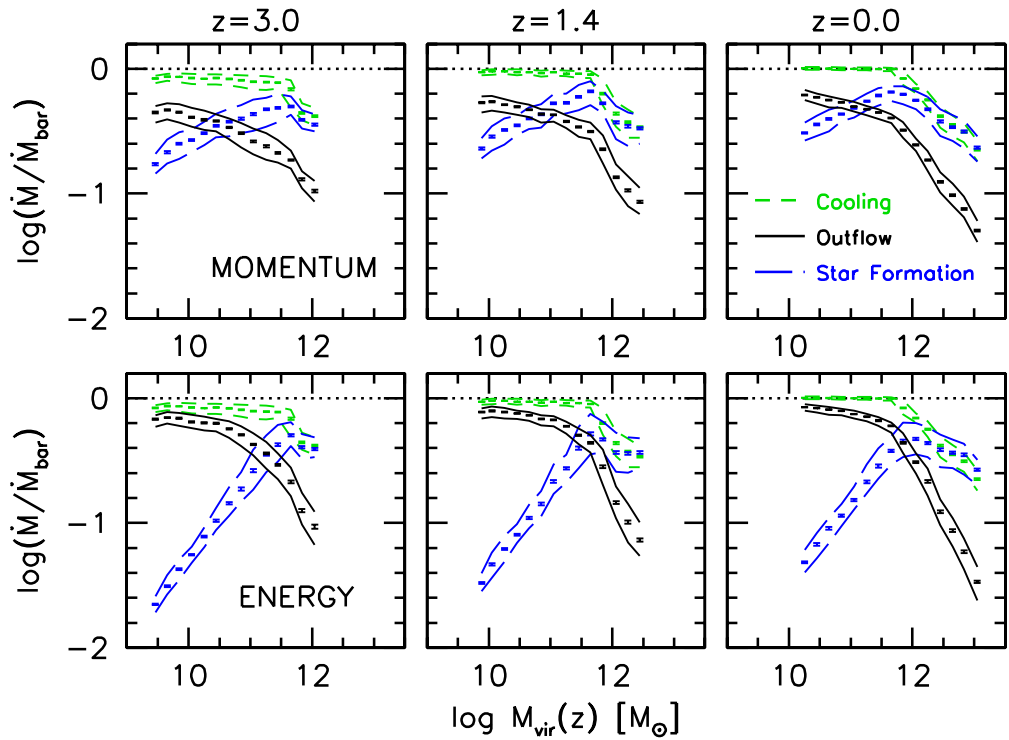,width=0.8\textwidth}}
\caption[]{\footnotesize Ratio between rate of cooling (green), star
  formation (blue) and outflow (black) to the rate of baryon accretion
  as a function of virial mass.  The upper panels are for momentum
  driven winds, the lower panels for energy driven winds. The panels
  from the left to right are for models at redshift, 3.0, 1.4, and
  0.0, respectively.  For haloes below $\simeq 1\times 10^{12} \Msun$,
  the cooling rate is essentially equal to the baryon accretion rate.
  The relative star formation rate increases with halo mass, whereas
  the relative outflow rate shows exactly the opposite trend. Thus
  globally the outflow rate is dominated by the depth of the potential
  well, rather than the efficiency of star formation.  Outflow rates
  are higher for energy driven winds due to the higher mass loading
  factor, this results in lower mass and surface density galaxies, and
  hence lower relative star formation rates.}
\label{fig:mdot3}
\end{figure*}

The middle panels of Fig.~\ref{fig:vw3} show the relation between wind
velocity and stellar mass. For  galaxies in the mass range $10^{9.6} <
\Mstar  <  10^{10.8} \Msun$  this  relation  has  a slope  of  $\simeq
0.33\pm0.01$ for all redshifts, and  both wind models.  For the energy
wind  model  this  relation  has  a shallower  slope  at  low  stellar
masses. The slope and small  scatter of this relation is a consequence
of the tight  correlation between $V_{\rm w}$ and  $V_{2.2I}$, and the
small     scatter    in     the    stellar     mass     TF    relation
(Figs.~\ref{fig:VMRId-mfb-nac} \&  \ref{fig:VMRId-efb-nac}).  The zero
point of the  relation between wind velocity and  stellar mass evolves
with  redshift, which is  a reflection  of lower  stellar masses  at a
given rotation  velocity in higher  redshift galaxies. Note  that this
implies evolution in  the zero point of the  stellar mass TF relation,
which will be discussed in a future paper.

The right panels of  Fig.~\ref{fig:vw3} show the relation between wind
velocity and star  formation rate.  This relation has  a slope $\simeq
0.38\pm0.01$ at $z=1.4$ for both  wind models. Galaxies at $z=1.4$ and
$z=3.0$ have similar  zero points, but galaxies at  $z=0.0$ have lower
star formation rates at a given wind velocity.

Over-plotted in red in Fig.~\ref{fig:vw3} are the observational
results at $z\simeq 1.4$ for the wind velocity versus stellar mass and
star formation rate from Weiner \etal (2009). These results are based
on a sample of 1406 galaxy spectra from the DEEP2 (DEEP Extragalactic
Evolutionary Probe 2) redshift survey (Davis \etal 2003) which have
both [O II] 3727 emission lines (to provide secure redshifts) and Mg
II $\lambda \lambda \,2796, 2803 \rm \AA$ absorption lines (to probe
low ionization outflowing gas). For the wind velocity we adopt the
velocity where Mg II absorption is at 10 \% of the continuum.

The slope of  the wind velocity stellar mass  relation is shallower in
the data  than our  models.  However, the  slope of the  wind velocity
star formation rate is  consistent. The differences between the slopes
of the  wind velocity - stellar  mass relation of the  models and data
may  be caused  by a  bias against  low mass  low star  formation rate
galaxies  in the  observations.  Martin (2005)  find  a similar
scaling between wind  velocity and SFR in low  redshift ultra luminous
infrared galaxies.  Thus both  energy and momentum wind models predict
scalings  that are  broadly  consistent with  the observations.   This
agreement is primarily due to our assumption that the wind velocity is
equal  to the local  escape velocity.   It is  surprising that  such a
naive  assumption,  which  is  most  likely  incorrect,  is  in  good
agreement with the observations.

We have shown that the  energy and momentum driven wind models predict
very similar scalings between wind  velocity and stellar mass and star
formation rate.  The slopes  of these relations are only significantly
different   at   low   masses    and   low   star   formation   rates.
Fig.~\ref{fig:etav2} shows  the mass  loading factor vs  wind velocity
for the momentum and energy driven wind models. The relations from our
galaxy formation models are independent  of redshift, and close to the
expected relations (shown as  red lines in Fig.~\ref{fig:etav2}) for a
model  where the wind  velocity is  independent of  radius. For  a wind
velocity of  $\simeq 100 \kms$ the  mass loading factor  of the energy
wind model is  an order of magnitude higher than  that of the momentum
wind  model. Thus  observationally, it  may be  easier  to distinguish
between energy and momentum driven winds by measuring the mass loading
factors of low mass galaxies.

\subsection{Relation Between Outflows, Inflows and Star Formation}
Fig.~\ref{fig:mdot3} shows  the ratio  between rate of  (cooling, star
formation, and outflow) to that of baryon accretion, versus halo mass,
at redshifts  $z=3.0$ (left),  $z=1.4$ (middle), and  $z=0.0$ (right).
In our model  the baryon accretion rate is  the cosmic baryon fraction
times the  halo accretion  rate, which is  a function of  the redshift
zero halo mass and concentration.   Thus higher mass halos have higher
rates of baryon accretion.

Cooling is efficient  in haloes below $\simeq 10^{12}  \Msun$, so that
the rate of  cooling is almost equal to the  rate of baryon accretion.
Above $\simeq 10^{12}\Msun$ the cooling efficiency drops, and thus the
rate  of  cooling  drops   significantly  below  the  rate  of  baryon
accretion.  The  scatter in the cooling  rate is caused  by scatter in
the  halo concentration,  which effects  the density  of the  halo and
hence temperature of the hot gas.
 
The star formation  rate, relative to the cooling  rate increases with
halo mass, with a much stronger dependence in energy driven winds than
momentum  driven  winds.    Above  $10^{12}\Msun$  the  relative  star
formation  rate decreases,  but  this is  due  to the  decline in  the
cooling rate,  rather than  inefficient star formation.   The relative
outflow rate shows the opposite  trend. Outflows are more efficient in
lower mass haloes,  this is despite the lower  relative star formation
rates.  For  haloes with  $\Mvir \lta 10^{12}  \Msun$ the  energy wind
model has higher  outflow rates than the momentum  wind model. This is
due to the higher mass loading  factor in the energy wind model, which
can be of order $\sim 10$ in low mass haloes.


\section{Summary}
\label{sec:sum}

We use a disk galaxy evolution model to investigate the impact of mass
outflows (a.k.a.   feedback) on disk galaxy  scaling relations, galaxy
mass fractions and spin  parameters.  Our model follows the accretion,
cooling,  star formation  and  ejection of  mass  inside growing  dark
matter   haloes  with   cosmologically   motivated  angular   momentum
distributions.   In our  models  the surface  density  profile of  the
baryonic  disk   is  determined  by  the   specific  angular  momentum
distribution  of   the  cooled  baryons  and  by   the  assumption  of
centrifugal equilibrium.   The surface density profile  of the stellar
disk is then  determined by the efficiency of  star formation with gas
density.   We model star  formation with  a Schmidt  law on  the dense
molecular gas.   We compute the molecular fraction  using the pressure
based prescription  in Blitz \&  Rosolowsky (2006).  We  consider both
energy and  momentum driven  galaxy wind models.   For both  models we
assume that  the wind velocity is  equal to the  local escape velocity
(from the disk and halo). This assumption maximizes the amount of mass
loss.  Our main conclusions are summarized as follows:

\begin{itemize}
\item  {\it Velocity-Mass-Radius Scaling  Relations: }  Models without
  feedback  result in disks  that, at  a given  stellar mass,  are too
  small   and   rotate    too   fast   (Figs.~\ref{fig:VMRIm-efb}   \&
  \ref{fig:VMRId-nfb-nac}).  With increasing feedback efficiency disks
  become larger,  and rotate slower  (Fig.~\ref{fig:VMRIm-efb}).  With
  high feedback  efficiency, disks  are too large  at a  given stellar
  mass.   However,   the  offset  from  the  TF   relation  is  almost
  independent of  feedback, because feedback reduces  both the stellar
  mass and rotation velocity (Fig.~\ref{fig:VMRIm-efb}).

\item  {\it Tully-Fisher Zero  Point: }  Models with  halo contraction
  over predict  the zero point of the  stellar mass-velocity relation,
  independent of the  feedback efficiency (Figs.~\ref{fig:VMRIm-efb} \&
  \ref{fig:VMRIm-mfb}).  Models  without halo contraction  result in a
  better agreement, but still  over predict the rotation velocities at
  high      stellar      masses      (Figs.~\ref{fig:VMRIm-efb}      \&
  \ref{fig:VMRIm-mfb}).

\item {\it The Baryonic Tully-Fisher Relation: } The Baryonic
  Tully-Fisher relation, defined as the relation between the (cold)
  baryonic mass and the circular velocity at large radius in the gas
  disk is only weakly sensitive to the feedback model
  (Fig.~\ref{fig:btf}). In particular, our energy driven feedback
  model (which results in galaxy mass fractions decreasing with
  decreasing halo mass) yields an almost identical slope to the no
  feedback model (which in which galaxy mass fractions are constant
  with halo mass).

\item {\it Gas fractions vs Stellar Mass: } Models without feedback
  predict gas fractions that are too low (Fig.~\ref{fig:fgas}), which
  is a result of the disks having densities that are too high, which
  in turn results in star formation being too efficient.  Models with
  feedback predict gas fractions in good agreement with observations
  (Fig.~\ref{fig:fgas}).

\item {\it Mass Fractions vs Halo Mass: } 
  Without feedback,  cooling is  very efficient below  a halo  mass of
  $\simeq 10^{12}\Msun$ (Fig.~\ref{fig:mglg3}).   Below this mass, the
  galaxy formation efficiency (defined  as the percentage of universal
  baryons that  end up as stars  and cold gas,  i.e. $\mgal/\fbar$) is
  constant at  $\simeq 95\%$, while above this  mass, galaxy formation
  efficiency decreases.   In both energy and  momentum feedback models
  mass is  more easily  ejected from lower  mass haloes,  resulting in
  galaxy  mass  fractions that  increase  with  halo  mass (below  the
  cooling threshold).  Maximum  galaxy formation efficiencies occur at
  a virial mass of $\Mvir \simeq 10^{12} \Msun$.  Maximum efficiencies
  as low  as $\simeq  35\%$ can be  produced with energy  driven winds
  with a feedback  efficiency of 0.25.  However, even  with a feedback
  efficiency  of 1,  momentum driven  winds result  in  maximum galaxy
  formation  efficiencies  of $\simeq  50\%$.  See  below for  further
  discussion.

\item {\it Mass Fractions vs Spin Parameter: } 
  At a  given halo  mass, higher density  disks are more  efficient at
  removing mass (Fig.~\ref{fig:dmg3}):  i.e.  energy/momentum input is
  more important than depth of potential, inconsistent with assumption
  of Gnedin \etal (2007).  At a given halo mass, star formation is more
  efficient  in  galaxies  with  higher surface  densities,  thus  gas
  fractions  are  lower.   This   helps  reduce  the  surface  density
  dependence of the stellar mass  TF relation relative to the baryonic
  TF relation (as argued by Firmani \& Avila-Reese 2000; van den Bosch
  2000; Dutton \etal 2007).

\item {\it Spin Parameter of Baryons vs. Dark Matter: }
  In haloes  with masses lower than $\simeq  10^{12}\Msun$, the galaxy
  spin  is  higher than  halo  spin,  because feedback  preferentially
  removes  low angular  momentum material  (Fig.~\ref{fig:mglg3}).  In
  haloes more  massive than $\simeq  10^{12}\Msun$ the galaxy  spin is
  lower than halo  spin, because that gas that has  not cooled has the
  highest specific angular momentum.  At a given halo mass the spin of
  the galaxy  is tightly  correlated with the  spin of the  halo, i.e.
  the  parameter  $(\jgal/\mgal)$  has   a  scatter  of  only  $\simeq
  0.02-0.04$ dex (Fig.~\ref{fig:jm3}).  This tight correlation is true
  even   for  galaxies   that   have  lost   over   $80\%$  of   their
  baryons. However,  due to the density dependence  of star formation,
  the  parameter $(\jstar/\mstar)$  has  a larger  scatter of  $\simeq
  0.08$ dex. See below for further discussion.

\item {\it Spin Parameter of Baryons vs. Stars: }
  The spin  of the stars is always  less than the spin  of the baryons
  because  star formation is  more efficient  at smaller  radii. Below
  $\Mvir \simeq  10^{12}\Msun$, the median spin parameter  of the stars
  is roughly independent of halo mass (Fig.~\ref{fig:mglg3}).

\item {\it Wind Velocity vs Galaxy Observables: } 
  Both feedback  models result in tight correlations  between the mean
  wind  velocity,  $V_{\rm  w}$,  stellar  mass,  $\Mstar$,  and  star
  formation  rate,  $\dot{M}_{\rm star}$:  $V_{\rm w}  \propto 3  V_{2.2I}
  \propto    \Mstar^{0.33}    \propto    \dot{M}_{\rm    star}^{0.38}$
  (Fig.~\ref{fig:vw3}).     The   first   relation    is   essentially
  construction,  since we  assume the  wind velocity  is equal  to the
  local escape  velocity from the galaxy-halo system.   The second and
  third relations are non-trivial.  The scaling relations between wind
  velocity and stellar mass/star formation rate are broadly consistent
  with  observations at  $z\sim 1.4$  from the  DEEP2  redshift survey
  (Weiner \etal 2009).

\item{\it Differences between Energy and Momentum Driven Winds:
  } 
  The main  difference is the  mass loading factor, which  scales like
  $V_{\rm w} ^{-2}$ for energy  driven winds, and $V_{\rm w}^{-1}$ for
  momentum  driven   winds.   For  equal  fractions   of  the  initial
  energy/momentum from  the SN that drives the  outflow, energy driven
  winds  have  higher  mass  loading  factors for  all  relevant  wind
  velocities.   The  differences  between  the  mass  loading  factors
  increase with decreasing  wind velocity (Fig.~\ref{fig:etav}).  Thus
  energy driven  winds are much  more efficient at removing  mass from
  lower mass  haloes (Fig.~\ref{fig:mglg3}).  This has  at least three
  observational consequences: 1)  different slopes of the size-stellar
  mass       relation       ($\simeq       0.14$      for       energy
  Fig.~\ref{fig:VMRId-efb-nac},   and  $\simeq  0.28$   for  momentum,
  Fig.~\ref{fig:VMRId-mfb-nac}); 2) different  slopes of the relations
  between  $\mgal$ and  $\mstar$ with  $\Mvir$ (see  Table 2);  and 3)
  different  slopes  in   the  metallicity-stellar  mass  relation  at
  $z\simeq  2$  ($\simeq 0.17$  for  momentum  and  $\simeq 0.33$  for
  energy, Fig.~\ref{fig:zm3}). See below for further discussion.

\item {\it Outflow vs Inflow: }
  The median mass outflow rate relative to the median mass inflow rate
  {\it decreases}  with increasing halo mass, whereas  the median star
  formation rate  relative to the  median inflow rate  {\it increases}
  with increasing  halo mass (Fig.~\ref{fig:mdot3}).   Thus, globally,
  the   depth  of   the   potential  is   more   important  than   the
  energy/momentum input from supernova.

\end{itemize}

\subsection{Comments on Galaxy Spin vs Halo Spin}

The assumption that $\lamgal=\lambda$ (i.e. $\jgal/\mgal=1$) underlies
almost all  analytical and  semi-analytical models  of disk  galaxy formation
(e.g.   Dalcanton, Spergel  \& Summers  1997; Mo,  Mao \&  White 1998;
Somerville \&  Primack 1999; Firmani  \& Avila-Reese 2000;  Cole \etal
2000; Croton  \etal 2006; Dutton  \etal 2007; Somerville  \etal 2008).
We have shown  that this assumption is no longer  valid in models with
outflows. A similar conclusion was reached by Maller \& Dekel (2002). 


The result  that $\lamgal$ is  significantly higher than  $\lambda$ in
low  mass haloes  helps to  resolve  the puzzle  surrounding the  spin
parameters  of bulge-less  dwarf galaxies.   Using observations  of 14
late-type  dwarf galaxies  van den  Bosch, Burkert  \&  Swaters (2001)
found the distribution of $\lamgal$ to have a median of $\simeq 0.06$.
D'Onghia \& Burkert (2004) measured  the spin parameter of dark matter
haloes  that are  most likely  to host  bulge-less disk  galaxies, and
found  a  median   spin  parameter  $\bar{\lambda}\simeq0.028$  (after
correcting  to  the energy  definition  of  halo  spin).  D'Onghia  \&
Burkert   (2004)   assumed   that    in   the   best   case   scenario
$\lamgal=\lambda$,  and thus  there is  a discrepancy  of a  factor of
$\simeq  2$  between  the  observed  and  predicted  spin  parameters.
However, with  energy driven feedback, our models  produce galaxy spin
parameters a factor  of 2 higher than the halo  spin parameters in low
mass haloes, thus resolving the discrepancy.

\subsection{Comments on Energy vs Momentum Driven Winds} 

In our models energy and momentum driven winds result in significantly
different slopes of the relations  between disk size and stellar mass.
Observations of the size-stellar  mass relation for late-type galaxies
from Shen  \etal (2003) find  a slope of  $0.14$ at the low  mass end,
which  favors our  energy wind  model  over our  momentum wind  model.
However, at the high mass end, a number of authors find steeper slopes
(e.g.   Shen \etal  2003; Pizagno  \etal 2005;  Courteau  \etal 2007),
which is in  better agreement with our momentum  wind model.  However,
there  are a  number of  uncertainties  in the  observations (such  as
determinations of stellar masses, inclination effects on galaxy sizes)
which need to be quantified before firmer conclusions can be made.

Our   energy  and  momentum   driven  wind   models  also   result  in
significantly different  slopes to the relations  between galaxy mass,
stellar  mass and  halo mass.   These  relations can  in principle  be
directly  tested  with  galaxy-galaxy  weak lensing  and/or  satellite
kinematics. 

\subsection{Comments on Why Galaxy Formation is Inefficient}

Observations of halo masses  from weak lensing studies (e.g.  Hoekstra
\etal 2005; Mandelbaum \etal 2006)  and methods that match the stellar
mass function to the halo  mass function (e.g. Yang \etal 2007; Conroy
\& Wechsler 2008)  find that the peak galaxy  formation efficiency has
to be  relatively low  $\simeq 0.33$.  We  have shown that  low galaxy
formation efficiencies are also required to explain the zero points of
the  relations  between  velocity,  stellar  mass, and  size  of  disk
galaxies.

We  have  shown that  mass  ejection  through  supernova driven  winds
provides  at  least  a  partial  explanation for  this.   However,  by
assuming that the wind velocity is equal to the local escape velocity,
results in  maximal outflow rates  for a given  energy/momentum input.
While such a  scaling of wind velocity with  galaxy escape velocity is
at least supported by observations  at low and high redshift, galactic
winds will  likely have a range  of velocities, which  will reduce the
outflow rates from  those in our model.  Needless  to say, the scaling
between wind velocity and escape velocity, as well as the mass loading
factor,   need  to   be  investigated   further   with  hydrodynamical
simulations. 

Even though we have adopted  a maximally efficient mass outflow model,
we still need  to use 25 percent  of all SN energy (or  100 percent of
all SN  momentum) in order  to eject enough  mass to fit the  data. It
remains to  be seen  whether such high  efficiencies are  realistic or
not. This requires detailed hydro-dynamical simulations with radiative
transfer, that accurately  model the complicated multi-phase structure
of  the  ISM.   It is  likely  that  one  needs to  invoke  additional
mechanisms  to explain  the low  baryonic mass  fractions  observed in
galaxy mass haloes.

An alternative  explanation for  low galaxy formation  efficiencies is
that  most of the  baryons never  accrete onto  galaxies in  the first
place.  In massive  haloes $\Mvir \gta 10^{12}\Msun$ (in  which gas is
heated  by an  accretion shock),  accretion onto  the galaxy  could be
suppressed  with  multi-phase  cooling  (Maller \&  Bullock  2004)  or
additional heating such as from  AGN, or mergers. However, in low mass
haloes, most of the baryons are accreted in cold streams.  Rather than
disrupting these streams from the  outside, such as with feedback from
the central  galaxy, a more likely  scenario would be  to disrupt them
from the  inside, i.e.  by  re-heating baryons and ejecting  them from
the cold  flow into the inter  galactic medium, {\it  before} the cold
flow  reaches the  halo.   We have  shown  that even  though the  star
formation  efficiency is  much lower  in lower  mass haloes,  the mass
loading factor is typically  high, especially for energy driven winds.
Thus even small amounts of star  formation in low mass haloes could be
sufficient  to significantly  reduce  the baryon  accretion rate,  and
hence baryon mass fraction of  the main galaxy.  Coupled with outflows
from the  main galaxy, this could  result in galaxy  mass fractions in
better  agreement  with  observations,  but  with a  lower  (and  more
realistic) conversion  efficiency of SN  energy/momentum into galactic
winds than required by our current models.

\section*{Acknowledgements} 

We thank, Eric Bell, Neil Katz, Andrea Macci\`o, Julio Navarro and
Padelis Papadopoulous for useful discussions, and the referee for
providing useful suggestions for improving the manuscript.  A.A.D.
acknowledges support from the National Science Foundation Grants
AST-0507483, AST-0808133, and the Swiss National Science
Foundation (SNF).

{\small

}

\label{lastpage}

\end{document}